\documentstyle[12pt,epsf]{article}
\def\journal#1, #2, #3, #4 { {\sl #1~}{\bf #2~}(#3) #4 }

\def\cmp{\journal Comm. Math. Phys., }

\def\np{\journal Nucl. Phys., }

\def\pl{\journal Phys. Lett., }

\catcode`\@=11
\def\marginnote#1{}
\newcount\hour
\newcount\minute
\newtoks\amorpm
\hour=\time\divide\hour by60
\minute=\time{\multiply\hour by60 \global\advance\minute
by-\hour}\edef\standardtime{{\ifnum\hour<12
\global\amorpm={am}%
        \else\global\amorpm={pm}\advance\hour by-12 \fi
        \ifnum\hour=0 \hour=12 \fi
        \number\hour:\ifnum\minute<10
0\fi\number\minute\the\amorpm}}
\edef\militarytime{\number\hour:\ifnum\minute<10
0\fi\number\minute}

\def\draftlabel#1{{\@bsphack\if@filesw {\let\thepage\relax
   \xdef\@gtempa{\write\@auxout{\string
      \newlabel{#1}{{\@currentlabel}{\thepage}}}}}\@gtempa
   \if@nobreak \ifvmode\nobreak\fi\fi\fi\@esphack}
        \gdef\@eqnlabel{#1}}
\def\@eqnlabel{}
\def\@vacuum{}
\def\draftmarginnote#1{\marginpar{\raggedright\scriptsize\tt#1}}
\def\draft{\oddsidemargin -.5truein
        \def\@oddfoot{\sl preliminary draft \hfil
        \rm\thepage\hfil\sl\today\quad\militarytime}
        \let\@evenfoot\@oddfoot \overfullrule 3pt
        \let\label=\draftlabel
        \let\marginnote=\draftmarginnote

\def\@eqnnum{(\theequation)\rlap{\kern\marginparsep\tt\@eqnlabel}%
\global\let\@eqnlabel\@vacuum}  }

\def\numberbysection{\@addtoreset{equation}{section}
        \def\theequation{\thesection.\arabic{equation}}}

\def\underline#1{\relax\ifmmode\@@underline#1\else
 $\@@underline{\hbox{#1}}$\relax\fi}

\catcode`@=12
\relax

\numberbysection
\def\souligne#1{\underline{#1}}
\def\fin{\end{document}}
\def\beq{\begin{equation}}
\def\eeq{\end{equation}}
\def\beqa{\begin{eqnarray}}
\def\eeqa{\end{eqnarray}}
 \def\nnn{\nonumber \\}
\def\sqr#1#2{{\vcenter{\vbox{\hrule height.#2pt
\hbox{\vrule width.#2pt height#1pt \kern#1pt
\vrule width.#2pt}
\hrule height.#2pt}}}}

\def\lfloorhat{{\hat \lfloor}}
\def\rfloorhat{{\hat \rfloor}}

\def\Je{J^e}
\def\Jeb{{\overline J}^e\, \!}
\def\Jep{J'\, \!^e\, \!}
\def\Jebp{{\overline J}'\, \! \!^e\, \!}
\def\Jne#1 {J_{#1}^e\, \!}
\def\Jneb#1 {{\overline J}_{#1}^e\, \!}
\def\Jnep#1 {J_{#1}'\, \!^e\, \!}
\def\Jnebp#1 {{\overline J}_{#1}'\, \!  \!^e\, \!}

\def\Jehat{{\widehat J}^e}
\def\hhat{{\widehat h}}

\def\Jhat{{\widehat J}}

\def\Chat{{\widehat C}}

\def\Nhat{{\widehat N}}

\def\mhat{{\widehat m}}

\def\Nhat{{\widehat N}}
\def\Vhat{{\widehat V}}

\def\gp{g' \, \!}
\def\gb{{\overline g}}
\def\gbp{{\overline g' \,\!}}

\def\chip{ \chi'\, \!}
\def\chib{{\overline \chi}}
\def\chibp{{\overline \chi}'\,\!}

\def\Vhat{{\widehat V}}

\def\rhat{{\widehat r}}

\def\qhat{{\widehat q}}
\def\varpip{\varpi'\,}
\def\varpibp{{\overline \varpi}'\,}
\def\varpihat{{\widehat \varpi}}
\def\varpihatp{{\widehat \varpi}'\,}
\def\varpihatb{{\widehat {\overline \varpi}}}
\def\varpihatbp{{\widehat {\overline \varpi}}'\,}
\def\varpib{{\overline \varpi}}

\def\jhat{{\widehat {\jmath}}}

\def\mhat{{\widehat m}}

\def\phat{{\widehat p}}
\def\Fhat{{\widehat F}}
\def\Shat{{\widehat S}}

\def\Jb{{\overline J}}

\def\nb{{\bar n}}
\def\rb{{\bar r}}

\def\Jge{{\underline J}}
\def\mge{{\underline m}}

\def\Jgen#1 {  {\underline J_{#1}} }
\def\Jgenp#1 #2 {(J_{#1}+{#2},\Jhat_{#1})}
\def\Jgenm#1 #2 {(J_{#1}-{#2},\Jhat_{#1})}
\def\Jg#1 {J_{#1},\Jhat_{#1}}
\def\Jgp#1 #2 {J_{#1}+{#2},\Jhat_{#1}}
\def\Mgen#1 {{\underline M_{#1}}}

\def\produit#1,#2,#3,#4 {P\Bigl ( [{#1},{#2}]\otimes\{{#3}\},{#4}\Bigr )}
\def\produitscript#1,#2,#3,#4 {P\Bigl (
[{\scriptstyle{#1},{#2}}]\otimes\{{\scriptstyle{#3}}\},{#4}\Bigr )}
\def\pprod#1,#2,#3,#4,#5 {P\Bigl ( [{#1},{#2}]\otimes[{#3},{#4}],{#5}\Bigr )}
\def\pprodscript#1,#2,#3,#4,#5 {P\Bigl (
[{\scriptstyle{#1},{#2}}]\otimes[{\scriptstyle{#3},{#4}}],{#5}\Bigr )}

\def\fusV#1,#2,#3,#4,#5,#6 {f_V(
\Jgen{#1} ,
\Jgen{#2} ,
\Jgen{#3} ,
\Jgen{#4} ,
\Jgen{#5} ,
\Jgen{#6} )}

\def\brdV#1,#2,#3,#4,#5,#6 {b_V(
\Jgen{#1} ,
\Jgen{#2} ,
\Jgen{#3} ,
\Jgen{#4} ,
\Jgen{#5} ,
\Jgen{#6} )}

\def\fusxi#1,#2,#3 {f_\xi (\Jgen{#1} ,
\Mgen{#1} ,
\Jgen{#2} ,
\Mgen{#2} ,
\Jgen{#3} )}

\def\gaghat{{\hat {\bigl \{}}}
\def\gadhat{{\hat {\bigr \}}}}

\def\bverthat{{\hat {\bigl \vert}}}

\def\sixjxi#1,#2,#3,#4,#5,#6 {{\left\{\left . \!\! \,^{#1}_{#2}
\,^{#3}_{#4} \right | \!\, ^{#5}_{#6}\right\}}}
\def\sixje#1,#2,#3,#4,#5,#6 {{\left\{\left\{\left . \!\! \, ^{#1}_{#2}
\, ^{#3}_{#4} \right | \!\, ^{#5}_{#6}\right\}\right\}}}
\def\sixjxihat#1,#2,#3,#4,#5,#6 {{{\gaghat\left . \!\! \, ^{#1}_{#2}
\, ^{#3}_{#4} \right | \!\, ^{#5}_{#6}\gadhat}}}
\def\sixjehat#1,#2,#3,#4,#5,#6 {{\gaghat\gaghat\left . \!\! \, ^{#1}_{#2}
\, ^{#3}_{#4} \right | \!\, ^{#5}_{#6}\gadhat\gadhat}}

\def\Jpe {J^{e+}}
\def\Jme {J^{e-}}
\def\Jpehat {\Jhat^{e+}}
\def\Jmehat {\Jhat^{e-}}
\def\Jehat {{\widehat J^e}}

\def\epsilonhat{{\widehat \epsilon}}
\def\pb{\overline p}
\def\phatb{\widehat {\overline p}}
\def\Jhatb{\widehat {\overline J}}

\begin{document}
\begin{titlepage}

\begin{flushright}

LPTENS--94/01, \\
hep-th@xxx/9403026, \\
February 1994
\end{flushright}

\vglue 1.5  true cm
\begin{center}
{\large \bf
SOLVING THE STRONGLY COUPLED 2D GRAVITY:\\
\medskip
2. FRACTIONAL-SPIN OPERATORS, AND \\
\medskip
TOPOLOGICAL THREE-POINT FUNCTIONS. } \\
\vglue 1.5 true cm
{\bf Jean-Loup~GERVAIS}\\
\medskip
{\bf Jean-Fran\c cois ROUSSEL}\\
\medskip
{\footnotesize Laboratoire de Physique Th\'eorique de
l'\'Ecole Normale Sup\'erieure\footnote{Unit\'e Propre du
Centre National de la Recherche Scientifique,
associ\'ee \`a l'\'Ecole Normale Sup\'erieure et \`a
l'Universit\'e
de Paris-Sud.},\\
24 rue Lhomond, 75231 Paris CEDEX 05, ~France}.
\end{center}
\vfill
\begin{abstract}
\baselineskip .4 true cm
\noindent
{\footnotesize
We report progress along the line  of
a previous article --- nb.  1 of the series ---
 by one of us  (J.-L. G.).
One main  point is   to include
chiral operators with fractional
quantum group spins (fourth or sixth
of integers) which are needed  to achieve the necessary
correspondence between
the set  of conformal weights of primaries
and
the physical spectrum of Virasoro highest weights.
This is possible by extending the  study   of the chiral
bootstrap (recently completed by
 E. Cremmer, and the present authors) to the case of
semi-infinite quantum-group representations which correspond to
positive integral screening numbers.
In particular, we prove the
Bidenharn-Elliot and Racah identities for
q-deformed 6-j symbols generalized to continuous spins.
The decoupling  of the
family of physical chiral operators (with real conformal weights)
at the special values $C_{Liouville}=7,$ $13$, and $19$,
is shown to provide
a  full solution of Moore and Seiberg's  equations, only
involving operators with real conformal weights.
Moreover, our study confirms the existence of
the strongly coupled topological models put forward earlier.
The three-point functions are determined.
 They are given by
a product of leg factors similar to the ones of the weakly coupled
models.
However, contrary to this latter case,
 the equality between the quantum group spins of the holomorphic
and antiholomorphic components is not preserved by the local
vertex operator. Thus   the
``c=1'' barrier appears as connected with a
deconfinement of chirality.
 }
\end{abstract}
\vfill
\end{titlepage}

\section{INTRODUCTION}
\markboth{ 1. Introduction }{ 1. Introduction }
At the present time, the only way\cite{GN,GR,G3} to go through
the ``$c=1$ barrier'' is to use
the operator approach to 2D gravity. The basic reason seems to be that one
should treat the two screening charges symmetrically in the strong coupling
regime, since they are complex conjugate. This is in sharp
contrast with what is
currently done in the weak coupling regime, say using matrix models.
In the operator approach, the quantum group structure was shown\cite{B,G1}
to be
of the type $U_q(sl(2))\odot U_\qhat (sl(2))$,
with $q=\exp ih$, $\qhat=\exp i\hhat$, and
$$
h={\pi \over 12}\Bigl(C-13 -
\sqrt {(C-25)(C-1)}\Bigr),
$$
\beq
\hhat={\pi \over 12}\Bigl(C-13
+\sqrt {(C-25)(C-1)}\Bigr),
\label{1.1}
\eeq
where $C$ is the central charge of Liouville theory.
Each quantum group parameter is associated with a screening charge by the
relations  $h=\pi (\alpha_-)^2/2$, $\hhat=\pi (\alpha_+)^2/2$.
In the strong coupling regime $1\leq C \leq 25$,
$h$ and $\hhat$ are complex conjugate.
Thus,  treating them symmetrically, as was done in refs.\cite{GN,GR,G3},
 is the key.
A major progress was made
in ref.\cite{G3}
 by proving a unitary truncation theorem, that holds  for special
values of $C$. The point of this theorem is as follows.
The basic family of $(r,s)$ chiral operators in 2D gravity may be labelled
by two quantum group spins $J$, and $\Jhat$, with
$r=2\Jhat+1$, $s=2J+1$, so that
the spectrum of
Virasoro weights is given by
\beq
 \Delta_{J\Jhat}
={C-1\over 24}-
{ 1 \over 24} \left((J+\Jhat+1) \sqrt{C-1}
-(J-\Jhat) \sqrt{C-25} \right)^2,
\label{1.2}
\eeq
in agreement with Kac's formula.
The  weak coupling regime corresponds to $C>25$ where
this formula is automatically real,
for real $J$, and $\Jhat$. In this paper,
we deal with  the strong coupling regime $1< C < 25$.
Then, since   $\sqrt{C-1}$ is real, and $\sqrt{C-25}$
pure imaginary, the  formula just written
 gives complex results in general.
However---in a way that is reminiscent
of the truncations that give the minimal
unitary models--- for $C=7$,
$13$, and $19$,  there is a consistent
truncation of the above general family
down to an operator algebra involving
operators with real Virasoro conformal
weights only. These  are of two types.
The first has spins $\Jhat=J$, and Virasoro weights that are negative;
the second has $\Jhat=-J-1$, and  Virasoro weights that are
positive.

In this article, we improve with respect to the
previous situation of ref.\cite{G3}  in several respects.
First, ref.\cite{G3} only dealt with the fusion algebra at the level
of primaries, making use of the quantum group
structure unravelled in refs.\cite{B,G1}.
The recent articles \cite{CGR1,CGR2}
which combine the operator
approach with the Moore and Seiberg scheme\cite{MS}, will  allow us  to deal
with
arbitrary descendants. Second, ref.\cite{G3} did not fully clarify the role
of the coupling constants  --- that are not governed solely by the
quantum group symmetry.	This may now be done using the
general formulae of refs.\cite{CGR1,CGR2}.
 Third, the spectrum of zero-modes of the physical
Hilbert space of the strongly coupled theories is such
that the corresponding
set
of operators   should involve in some cases chiral primary fields with
quantum group spins that are rational numbers instead of   halves  of
integers.   These operators were not considered in ref.\cite{G3}, since the
stucture of the chiral operator algebra was not yet known well enough to
generalize to non-half-integer spins.  This is now possible thanks to
ref.\cite{CGR1,CGR2}. Moreover, a recent
study\cite{GS1}\cite{GS3}  based on
an operator Coulomb-gas realization makes it possible to derive the
braiding of chiral vertex operators with arbitrary spins
provided the screening numbers remain integer.
Finally, the Liouville string associated with the strongly coupled theories
under consideration, remarkable as
they may be\cite{BG}, remain complicated.
It was proposed earlier\cite{G3}
 to consider strongly coupled topological models,
obtained by considering together two copies of Liouville, so that
the total central charge is $26$. Our study confirms the consistency of
these models. Using our determination of the coupling constants, we may
calculate their  three-point functions,
it is found to be a product of
leg factors much like the result of
matrix models for $c\leq 1$ as recovered
from the continuous Liouville theory in the present framework\cite{G5}.
However, a strikingly novel feature appears. In the strong coupling regime,
the Liouville exponentials  cannot be used, since they
 do not preserve the
physical Hilbert space. They are replaced\cite{G3}  by other
local vertex operators  which
 shift the zero modes of the left- and right-movers
in  uncorrelated ways, so that it is not consistent to assume
that their  quantum group spins $J$ and $\Jb$ are equal.
This is in sharp contrast with the Liouville field. It shows that
the strong coupling regime may be characterized by a
deconfinement of chirality.
\section{GETTING STARTED}
\markboth{ 2. Getting started }{ 2. Getting started }
In this section we go through some background material as a preparation
for the main body of the paper.
 All notations are the same as in the previous articles
using the operator method. We shall not fully re-explain the conventions.
One outcome of ref.\cite{CGR1} was the fusion and braiding of
the general chiral operators $V_{\mge}^{(\Jge )}$, also denoted
$V_{m \mhat}^{(J \Jhat)}$, where underlined  symbols denote double indices
$\Jge \equiv (J,  \, \Jhat)$, $\mge \equiv (m,\,  \mhat)$,  which were
all taken to be half-integers:
$$
{\cal P}_{\Jgen{} } V_{\Jgen23 -\Jgen{} }
^{(\Jgen1 )}
V_{\Jgen3 -\Jgen23 }^{(\Jgen2 )} =
\sum_{\Jgen12 }
{g_{\Jgen1 \Jgen2 }^{\Jgen{12} }\
g_{\Jgen{12} \Jgen3 }^{\Jgen{} }
\over
g _{\Jgen2 \Jgen3 }^{\Jgen{23} }\
g_{{\Jgen1 }\Jgen{23} }^{\Jgen{} }
}
\left\{
^{\Jgen1 }_{\Jgen3 }\,^{\Jgen2 }_{\Jgen{} }
\right. \left |^{\Jgen{12} }_{\Jgen{23} }\right\} \times
$$
\beq
{\cal P}_{\Jgen{} } \sum_{\{\nu\}}
V_{\Jgen3 -\Jgen{} } ^{(\Jgen12 ,\{\nu\} )}
<\!\varpi _\Jgen12 ,\{\nu\}  \vert
V ^{(\Jgen1 )}_{\Jgen2 -\Jgen12 } \vert \varpi_{\Jgen2 }\! >,
\label{2.1}
\eeq
$$
{\cal P}_{\Jgen{} } V_{\Jgen23 -\Jgen{} }
^{(\Jgen1 )}
V_{\Jgen3 -\Jgen23 }^{(\Jgen2 )} =
\sum_{\Jgen13 }
e^{\pm i\pi (\Delta_\Jgen{} +\Delta_\Jgen3
-\Delta_{\Jgen23 }-\Delta_{\Jgen13 })}\times
$$
\beq
{g_{\Jgen1 \Jgen3 }^{\Jgen{13} }\
g_{\Jgen{13} \Jgen3 }^{\Jgen{} }
\over
g _{\Jgen2 \Jgen3 }^{\Jgen{23} }\
g_{{\Jgen1 }\Jgen{23} }^{\Jgen{} }
}
\left\{
^{\Jgen1 }_{\Jgen2 }\,^{\Jgen3 }_{\Jgen{} }
\right. \left |^{\Jgen{12} }_{\Jgen{23} }\right\}
{\cal P}_{\Jgen{} }
V_{\Jgen13 -\Jgen{} } ^{(\Jgen2 )}
V_{\Jgen3 -\Jgen13 }^{(\Jgen1 )}.
\label{2.2}
\eeq
In these  formulae,  world-sheet variables are omitted,
and $\varpi$ is the rescaled zero-mode of the underlying
B\"acklund free field that characterizes the Verma modules
${\cal H}(\varpi)$ spanned  by states noted $|\varpi, \, \{\nu\}>$, where
$\{\nu\}$ is a multi-index. The symbol $\varpi_{\Jge}$ stands for
$\varpi_0+2J+2\Jhat \pi/h$ where $\varpi_0 =1+\pi/h$
corresponds to the $sl(2)$-invariant
vacuum; ${\cal P}_{\Jgen{} }$ is the projector on
${\cal H}(\varpi_{\Jge})$. The above formulae contain the recoupling
coefficients for the
quantum group structure $U_q(sl(2))\odot U_\qhat(sl(2))$, which are defined by
\beq
\left\{
^{\Jgen1 }_{\Jgen3 }\,^{\Jgen2 }_{\Jgen{} }
\right. \left |^{\Jgen{12} }_{\Jgen{23} }\right\}
=
(-1)^{ \fusV 1,2,,23,3,12 }
\left\{
^{J_1}_{J_3}\,^{J_2}_{J}
\right. \left |^{J_{12}}_{J_{23}}\right\}
\gaghat
\,^{\Jhat_1}_{\Jhat_3}\,^{\Jhat_2}_{\Jhat}
\bigr. \bverthat \, ^{\Jhat_{12}}_{\Jhat_{23}}\gadhat
\label{2.3}
\eeq
where $\fusV 1,2,,23,3,12 $ is an integer to which we shall come back.
The symbol  $\left\{
^{J_1}_{J_3}\,^{J_2}_{J}
\right. \left |^{J_{12}}_{J_{23}}\right\} $ is the 6-j coefficient
associated with  $U_q(sl(2))$, while $\gaghat
\,^{\Jhat_1}_{\Jhat_3}\,^{\Jhat_2}_{\Jhat}
\bigr. \bverthat \, ^{\Jhat_{12}}_{\Jhat_{23}}\gadhat$ stands for the
6-j  associated with $U_{\qhat}(sl(2))$. In addition to these group
theoretic features  there appear the coupling constants
	$g_{{\Jgen{1} }\Jgen{2} }^{\Jgen{12} }$ whose general expression
was derived earlier.

On the other hand, the following notions were introduced in the previous
work on strongly coupled Liouville theory\cite{GN,G3}.

\smallskip

\noindent \souligne{a) The physical Hilbert space.} It is given
by
\beq
{\cal H}_{s \,  \hbox {\scriptsize phys}}
\equiv  \bigoplus_{r=0}^{1-s}\,\bigoplus_{n=-\infty}^{_\infty}\>
{\cal H}_s\bigl(\varpi_{r,\, n}\bigr),
\label{2.5}
\eeq
\beq
\varpi_{r,\, n}
\equiv \Bigl({r\over 2-s}+n\Bigr)
\,\bigl(1-{\pi\over h}\bigr).
\label{2.6}
\eeq
 The integer $s$ is such that the special values correspond to
\beq
 C=1+6(s+2),\quad s=0,\,\pm 1,\quad h+\hhat=s \pi.
\label{2.7}
\eeq
The weight
$\Delta (\varpi_{r,\, n})
\equiv (1+\pi/h)^2h/4\pi -h \varpi_{r,\, n}^2/4\pi$ is positive and
 in ${\cal H}_{s \,  \hbox {\scriptsize phys}}$
the representation of the Virasoro algebra
  is unitary. In Eq.\ref{2.5} we added a subscript $s$
to indicate  that the Hilbert spaces
depend upon the central charge, so that they are explicit functions of $s$.
This will be useful in section 7.
The  partition
function corresponds to compactification on a circle
with radius $R=\sqrt {2(2-s)}$ (see refs.\cite{BG,GR}).
\smallskip

\noindent\souligne{b) The restricted set of conformal weights}. The
truncated family only involves
 operators of the type   $(2J+1,2J+1)$ noted $\chi_-^{(J)}$
and $(-2J-1, 2J+1)$ noted $\chi_+^{(J)}$. Their Virasoro
conformal weights\cite{GN,GR,G3} which are
respectively given by
\beq
\Delta^-(J, C)=-{C-1\over 6}\,J(J+1), \quad
\Delta^+(J, C)=1+{25-C\over 6}\, J(J+1),
\label{2.8}
\eeq
are real.  $\Delta^-(J)$ is negative for all $J$ (except for
$J=-1/2$
where it becomes equal to $\Delta^+(-1/2)=(s+2)/4$). $\Delta^+(J)$
is always positive, and is larger than one if $J \not= -1/2$.\smallskip

\noindent\souligne{c) The truncated families}:
${\cal A}^\pm_{phys}$
is the   set of operators noted $\chi_\pm^{(J)}$,
introduced in \cite{GN,GR,G3}, whose conformal weights are given by
Eq.\ref{2.8}. In ref.\cite{G3},
the case of integer $2J$ was completely solved at the level of
primaries. They were expressed as specific linear combinations of
 $V_{M,\, -M}^{(J,\, J)}$
(resp $V_{M,\, -M}^{(-J-1,\, J)}$), so that
the following holds.
\smallskip

\noindent {THE UNITARY TRUNCATION THEOREM}:

For $C=1+6(s+2)$, $s=0$, $\pm 1$,
and when it
acts on ${\cal H}_{s \,  \hbox {\scriptsize phys}}$;
the  set ${\cal A}^+_{phys}$
(resp. ${\cal A}^-_{phys}$)
of operators $\chi_{+}^{(J)}$ (resp. $\chi_{-}^{(J)}$)
 is closed by fusion and braiding, and
  only gives states that belong to ${\cal H}_{phys}$.

With the formulae just recalled we may explain the main point of our paper.
The fusion and braiding relations Eqs.\ref{2.1} and \ref{2.2} are operator
relations computed between states ${\cal H}(\varpi_{\Jge})$ on the left and
${\cal H}(\varpi_{\Jge_3})$ on the right. Yet all spins are treated
on the same footing, following the basic scheme of
Moore and Seiberg, where there is a one-to-one correspondence between
the spectra of highest weights  and conformal weights of
  primary fields.  This will force us to
an extensive generalization of the work of ref.\cite{CGR1,CGR2}.  First, in
Eqs.\ref{2.1}, \ref{2.2},
the spins $2J$, $2\Jhat$, $2J_3$, and $2\Jhat_3$  were
assumed to be integers. On the other hand, we may write Eq.\ref{2.6} as
\beq
\varpi_{r,n}=\varpi_{\Jge_{r,n}},
\quad  J_{r,n} = {r\over 2(2-s)}+{n-1\over 2},
\quad \Jhat_{r,n} = -{r\over 2(2-s)}-{n+1\over 2}
\label{2.15}.
\eeq
One sees that, for $s=-1$, and $s=0$, this introduces spins that are
rational numbers, but not halves of integers. Once we go away from
half-integer spins, we may as well consider continuous spins as is done in
refs.\cite{GS1,GS3}.
Second, the treatment of ref.\cite{G3} had another basic difference between
the spectrum of highest-weight states
 and of conformal  weights of primary operators: the physical spectrum of the
latter involves two types  of fields (${\cal A}^\pm_{phys}$), with
$\Jhat=J$, and $\Jhat=-J-1$, respectively, while the physical
Hilbert space only involved spins satisfying the latter type condition,
as is clear from Eq.\ref{2.15}. The reason is that the physical
Hilbert space was taken to have positive highest-weights  so that the
Virasoro  representation be unitary. From the viewpoint of the operator
algebra we are led to enlarge the Hilbert space and consider
\beq
{\cal H}^\pm_{s \,  \hbox {\scriptsize phys}}
\equiv  \bigoplus_{r=0}^{1\mp s}\,\bigoplus_{n=-\infty}^{_\infty}\>
{\cal H}_s\bigl(\varpi^\pm _{r,\, n}\bigr),
\label{2.16}
\eeq
\beq
\varpi^\pm_{r,\, n}
\equiv \Bigl({r\over 2\mp s}+n\Bigr)
\,\bigl(1\mp {\pi\over h}\bigr).
\label{2.17}
\eeq
Now Eq.\ref{2.9} is generalized to
\beq
\varpi^\pm _{r,n}=\varpi_{\Jge^\pm_{r,n}},
\quad  J^\pm_{r,n} = {r\over 2(2\mp s)}+{n-1\over 2},
\quad \Jhat^\pm_{r,n} = \mp {r\over 2(2 \mp s)}+{\mp n-1\over 2}
\label{2.18}.
\eeq
One sees that for $s=0$, the $J$'s may be fourths  of integers, while,
for $s=1$ (resp. $s=-1$) the $J^-$'s (resp. the $J^+$'s)  may be
sixths of integers.
A parenthetical remark is in order at this point. In general, a
highest-weight state $|\varpi>$ is eigenstate of $L_0$ with an eigenvalue
$(1+\pi/h)^2
h/4\pi-h\varpi^2/4\pi$ that is invariant under $\varpi\to -\varpi$.
Since the corresponding Verma module ${\cal H}(\varpi)$  may be deduced
group theoretically once this eigenvalue is fixed, it follows that
${\cal H}(\varpi)$ and ${\cal H}(-\varpi)$ are the same Hilbert space.
Changing
$\varpi_{r,n} \to -\varpi_{r,n}$ in Eq.\ref{2.17} is equivalent to
changing $J^\pm_{r,n}\to -J^\pm_{r,n}-1$, and $\Jhat^\pm_{r,n}
\to -\Jhat^\pm_{r,n}-1$. It is thus related to the symmetry put forward
in ref.\cite{G3}. We shall make use of this freedom below.
Returning to our main line, let us note that,
in ${\cal H}^-_{s phys}$,
the highest weights are real but negative.
What is its physical meaning? This brings in the proposal of ref.\cite{G3}.
There it was remarked that the spectrum of physical conformal weights
Eq.\ref{2.8}  is
such that
\beq
\Delta^{\pm}(J, C)+\Delta^\mp (J, 26-C)=1
\label{2.19}.
\eeq
Moreover, if $C$ takes the special values Eq.\ref{2.7} one has
\beq
26-C =1+6(s'+2),\quad s'=-s
\label{2.20}.
\eeq
Thus we may construct a consistent string model with two copies of
strongly coupled Liouville theories, one with $s$ playing the
role of gravity, and the other with $s'$ being considered as the  matter.
Following the usual counting where the BRST cohomology  removes  two
degrees of freedom,  this theory has
essentially no degree of freedom, meaning that it is topological.
This is why the Hilbert space ${\cal H}^-_{phys}$ may be used:
the excitations that would be negative-normed states decouple.
This situation is of course similar to the one of
the $c\leq 1$ topological models\footnote{The central charge of matter is
noted $c$.}. Thus we shall follow closely the
Liouville derivation\cite{G5}
 of the three-point function for the $c\leq 1$ models.

The plan of the article is as follows. In section 3
we first prove some mathematical properties concerning
the extension of the 6-j symbols
to non-integer spins. These  are
necessary to generalize the operator algebra to
non-integer spins, as done  in section 4. This  extension to non-integer spins
will be shown to consistently include representations in a generalized
sense that are semi-infinite. In the language of  the
Coulomb-gas approach to conformal theories, they correspond to chiral fields
with numbers  of screening operators that are positive integers.
This is  actually how they are explicitly constructed in refs.\cite{GS1,GS2}.
As such, they are not what is needed to define the operators of
the family  ${\cal A}^+_{phys}$, since these involve negative spins
and negative screening numbers.
In section 5 we thus  continue
our  results  to negative ``screening
numbers''
thanks to the continuation $J_i\to-J_i-1$, put forward
in ref.\cite{G3}, which we apply  to the generalized
6-j symbols. This involves
 a non trivial identity on q-deformed hypergeometric functions $_4F_3$.
We are then ready to build
the physical consistent algebra
(in section 6), and show that it is fully closed
by  fusion and braiding to all order in the secondaries.
In other words, at the special values we have a complete truncation of the
chiral bootstrap equations down to the one describing the OPA of a set
of  chiral fields with real conformal weights.
Finally, in section 7, we apply the formulae just derived  to compute
the three-point function of the  topological models mentioned  above.

Our discussion will follow the line of ref.\cite{G3}, by
establishing relations between the quantum group symbols of
$U_q(sl(2))$,
and $U_{\qhat}(sl(2))$, that hold at the special values, and ensure
that  the truncation theorem holds. However, we will be led
to change
the definition of the $\chi$ fields. Let us comment about this now.
In ref.\cite{G3}, the following expression for the
$\chi_-^{(J)}$ fields was used\footnote{It is distinguished by the subscript
``G''.}
\beq
\chi_{- \hbox{\scriptsize  G.}}^{(J)} =\sum_{m=-J}^J C_m^{(J)}(\varpi)
e^{im [\hhat \varpihat -h \varpi+(h-\hhat)/2]} \psi_{m,\, -m}^{(J, J)}
\label{2.9}
\eeq
\beq
C_m^{(J)}(\varpi) \propto (-1)^{J-m}
{2J \choose J-m} { \prod _{t=0} ^{2J} \lfloor \varpi-J+m-t\rfloor
\over \lfloor \varpi+2m\rfloor},
\label{2.10}
\eeq
where $\propto$ means that we work up to factors that only depend upon $J$.
The chiral fields $\psi_{m,\, -m}^{(J, J)}$ differ from
$V_{m,\, -m}^{(J, J)}$ by normalization factors,
 noted $E_{m,\, -m}^{(J, J)}$ in the recent papers. In ref.\cite{G3}, the
method used was as follows. One starts from the ansatz for the simplest
case
$$
\chi_-^{(1/2)} =a(\varpi) \psi _{-1/2,\, 1/2}^{(1/2, 1/2)}
+b(\varpi) \psi _{1/2,\, -1/2}^{(1/2, 1/2)}.
$$
Closure by braiding requires
\beq
a(\varpi_{rn})b(\varpi_{r n+1}) =\lambda \sin (h\varpi_{rn})
\sin (h\varpi_{r n+1})
\label{2.14}
\eeq
where  $\lambda$ is an arbitrary constant. The following  solution of
these equations was used
in refs.\cite{GR,G3}:
$$
a=e^{i\hhat \varpihat-h \varpi} e^{ih/2} \sin (h\varpi), \quad
b=-e^{i\hhat \varpihat-h \varpi} e^{i\hhat/2} \sin (h\varpi),
$$
which,  using leading-order fusion,  leads to Eq.\ref{2.9}.
This particular choice  was made
in order to arrive at simple expressions  in terms of the
quantum-group covariant chiral fields $\xi_{M,\, -M}^{(J, J)}$.
At that time, the complete braiding matrix was only known for the
$\xi$ fields, so that the discussion made an extensive use of them.
The present viewpoint is somewhat different.
We shall use the
$V$ fields instead of the $\psi$'s, or the $\xi$'s.
Moreover, we consider fusion to all
orders in the descendants. The   definitions  which appear natural
from the present standpoint look
 rather different from the ones of
ref.\cite{G3}.   It is thus useful to sketch how
Eq.\ref{2.9} is related with the general expression we will
derive
in section 6. This is done in an appendix.

\section{GENERALIZATION OF 6-J SYMBOLS}
\markboth{ 3. Generalization of 6-j's }{ 3. Generalization of 6-j's }

\label{s2}

In this section we propose a generalization of the 6-j symbols
to non-half-integer spins,
and prove the corresponding generalized polynomial equations\footnote{
This is an abusive use of the  name
``polynomial equations'' which usually refers to
consistency equations for  fusion
and braiding coefficients.
But the 6-j coefficients, which  are  solutions  of them,
satisfy parallel equations,
namely orthogonality, Racah identity,
Bidenharn-Elliot identity...,  that we generically
call polynomial equations as well.}
in particular the pentagonal relation.
Although purely mathematical,
this generalization stems from the needs of physics,
and we shall therefore follow this guide, in two steps.

The standard three-leg vertex operators
intertwine three standard representations
of $U_q(sl(2))$ labelled by positive\footnote{All along this
section, we use positive for non-negative, including zero.}
half-integers.
Their algebra was completely elucidated in ref.\cite{CGR1},
using Moore and Seiberg formalism.
However, since the early eighties Gervais and Neveu have
introduced operators of positive half-integer spins, but
acting on a Hilbert space described by a continuous
zero-mode $\varpi$ or equivalently,
by a continuous spin.
This will lead to the first step of generalization.
The discrete values $\varpi=\varpi_J=\varpi_0+2J$,
with $2J\in Z_+$,
give back the standard case.

The second step will come from
operators with three continuous spins.
Their braiding has already been elucidated in refs.\cite{GS1,GS3}
using a Coulomb-gas approach.
It is essentially given by the generalized 6-j introduced
by Askey and Wilson\cite{AW} who proved the orthogonality relation.
Our more generalized 6-j coincide with theirs, and moreover we prove
the other polynomial equations.

These two generalizations could seem to be of a very different kind,
but they actually follow one another very naturally.
The right language is not the one of half-integer positive
or continuous spins but a question of number of restrictions on the spins,
the former being only a consequence of the latter.
Let us be more explicit.
We use dotted vertices, a dot on one leg meaning that
the sum of the spins of the two other legs minus the spin of this leg
is constrained to be a positive integer. Since there are three legs
at a vertex we may have one, two or three dots. We call them
type TI1, TI2, TI3, where  the letters
T and I stand for triangular inequalities.
We illustrate the meaning of the dots:
$$
\epsffile{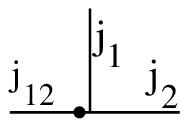}
\ \ \to\ \
j_1+j_2-j_{12}
\hbox{  positive integer}
$$
$$
\epsffile{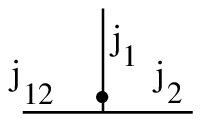}
\ \ \to\ \
j_2+j_{12}-j_1
\hbox{  positive integer}\footnote{
This vertex does not exist by itself as the only ones allowed are
\ref{ThreeCond}, \ref{TwoCond}, \ref{OneCond}.
It is only represented in order to illustrate
the dot convention.}.
$$
Adding dots to a vertex adds other restrictions.
For the type TI3, we have
\beq
\epsffile{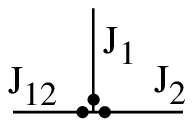}
\to\!
\left\{
\begin{array}{ccc}
J_1+J_2-J_{12} && \!\!\!\hbox{positive integer}\\
J_{12}+J_2-J_1 && \!\!\!\hbox{positive integer}\\
J_1+J_{12}-J_2 && \!\!\!\hbox{positive integer}\\
\end{array}
\right\}
\!\!\Rightarrow
\begin{array}{c}
2J_1,2J_2,2J_{12} \nnn
\hbox{ positive integers}.
\end{array}
\label{ThreeCond}
\eeq
These are but the usual (full) triangular inequalities,
or the branching rules for $sl(2)$, with half-integer spins.
So, the standard operators  are
of the   TI3 type.

The first step of generalization is to relax one restriction, which gives
the TI2 type:
\beq
\epsffile{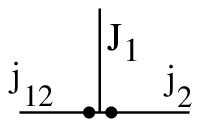}
\ \to\
\left\{
\begin{array}{ccc}
J_1+j_2-j_{12} \hbox{ positive integer}\\
J_1+j_{12}-j_2 \hbox{ positive integer}\\
\end{array}
\right\}
\ \Rightarrow\
2J_1\hbox{ positive integer}
\label{TwoCond}
\eeq
$j_{12}$ and $j_2$ being arbitrary\footnote{For the time being, we denote
half-integer positive spins by capital letters and
continuous  ones by small letters, but this is only a consequence
of the type of vertex and in no case
an a priori assumption.}.
In this TI2 case,  $J_1$ is a positive half-integer, and
$j_2-j_{12}$ is a half-integer (positive or negative).
This is fixed by  the number of restrictions.
The second step is of course to keep only one restriction (type TI1):
\beq
\epsffile{OneCond.ps}
\ \to\
j_1+j_2-j_{12} \hbox{ positive integer}
\label{OneCond}
\eeq
and none of the spins are half-integers.
Then the fusion or braiding of such generalized vertices lead
to 6-j symbols generalized in a very specific way,
and the ``miracle'' is that it is mathematically consistent.
In the first step of generalization,
fusion and braiding lead to two different
generalized 6-j, whereas they lead to only one kind of 6-j
in the second step.

\subsection{The first step of  generalization}

First, we try to define a 6-j coefficient for the following
fusion operation:
\beq
\epsffile{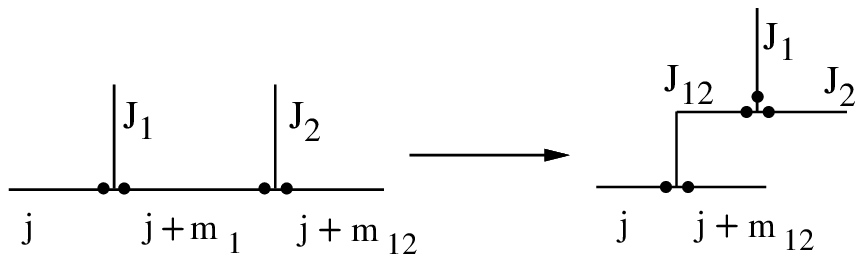}.
\label{FusTwoCond}
\eeq
It follows from the general rules for these three TI2 and one TI3
vertices that
$J_1$, $J_2$, $J_{12}$ are positive half-integers, that
$J_i\pm m_i$, for $i=1,2,12$
are positive integers ($m_{12}\equiv m_1+m_2$),
and that
$J_1$, $J_2$, $J_{12}$ satisfy the full triangular inequality.

The standard\cite{KR,HHM}
 (i.e. four TI3) definition of the q-deformed 6-j symbols
that is to be generalized, is:
$$
\left\{ ^{\quad J_1}_{j+m_{12}}\,^{J_2}_j\right | \left .
^{J_{12}}_{j+m_1}\right\}=
\sqrt{
\lfloor 2J_{12}+1 \rfloor
\lfloor 2j+2m_1+1 \rfloor
}
e^{i\pi (J_1+J_2-2J_{12}-m_{12}-2j)}
\times
$$
$$
\Delta(J_1,J_2,J_{12})\Delta(J_1,j,j+m_1)\Delta(J_{12},j,j+m_{12})
\Delta(J_2,j+m_1,j+m_{12})\times
$$
$$
\sum_{z}
{
e^{i\pi z}
\lfloor z+1 \rfloor \! !
\over
\lfloor z\!-\!J_1\!-\!J_2\!-\!J_{12} \rfloor \! !
\lfloor z\!-\!2j\!-\!J_1\!-\!m_1 \rfloor \! !
\lfloor z\!-\!2j\!-\!J_2\!-\!m_1\!-\!m_{12} \rfloor \! !
\lfloor z\!-\!2j\!-\!J_{12}\!-\!m_{12} \rfloor \! !
}
$$
\beq
{
1
\over
\lfloor J_1+J_2+m_{12}+2j-z \rfloor \! !
\lfloor J_1+J_{12}+m_1+m_{12}+2j-z \rfloor \! !
\lfloor J_2+J_{12}+m_1+2j-z \rfloor \! !
}
\label{sixj1}
\eeq
with
\beq
\Delta(l,j,k)=
\sqrt{\lfloor -l+j+k \rfloor \! !
\lfloor l-j+k \rfloor \! !
\lfloor l+j-k \rfloor \! !
\over
\lfloor l+j+k+1 \rfloor \! !}
\label{delta}
\eeq
and
\beq
\lfloor n \rfloor \! ! \equiv
\prod_{r=1}^n \lfloor r \rfloor,
\qquad \lfloor x \rfloor \equiv {\sin (hx)\over \sin h}.
\eeq

These 6-j coefficients involve square roots with
sign  ambiguities.
In the building of strong coupling regime operators,
the cancellation of signs is of a great importance (see section \ref{s5}),
and we must therefore treat them consistently. Following   ref.\cite{G5},
and since the basic mathematical tool is the relationship between
sine and gamma functions,
we define square roots of sine  functions for non-integer argument
  from the relation
$\sqrt{\sin (\pi z)}\equiv\sqrt{\pi}/
(\sqrt{\Gamma(z)}\sqrt{\Gamma(1-z)})$.
The choice of the sheet for $\sqrt{\Gamma(z)}$ is determined
so that the relation $\sqrt{\Gamma(z+1)}=\sqrt{z}\sqrt{\Gamma(z)}$
holds, where $\sqrt{z}$ is defined as usual
with a cut on the negative real axis.
This specifies
the relation between $\sqrt{\sin (\pi (z+1))}$ and
$\sqrt{\sin (\pi (z))}$. In the present case, one has
typically $z\sim (J_i+...)h/\pi$. If h is complex, as it is
the case in the strong coupling regime,
 $z$ is far from the cut. For the weak coupling regime, on the contrary,
$h$ is real, and one should temporarily give a small imaginary part
to $J$ or to $h$ to specify the sheet.

For simplicity we lump  the square roots
of 6-j symbols into  coefficients noted  $\Xi$.
Those
$\Xi$ factors are chosen so that they can be
seen as normalization factors of the  vertices
of Eq.\ref{FusTwoCond}, and will cancel (or factorize) out
of the polynomial equations when applying successive
fusions (or braidings).
So, the polynomial equations are fundamentally
rational equations (without square roots).
We could take $\Xi=\Delta$,
but $\Delta$ involves factorials with arguments which
will not remain integers in the forthcoming generalization
to non-half-integer spins.
This is why we define ($p_{1,2}\equiv j_1+j_2-j_{12}$)
$$
\Xi_{j_1j_2}^{j_{12}}\equiv
\prod_{k=1}^{p_{1,2}}
\sqrt{
\lfloor 2j_1-k+1 \rfloor
\lfloor 2j_2-k+1 \rfloor
\lfloor 2j_{12}+k+1 \rfloor
\over
\lfloor k \rfloor
}
$$
\beq
=
\sqrt{
\lfloor j_1-j_2+j_{12}+1 \rfloor_{p_{1,2}}
\lfloor -j_1+j_2+j_{12}+1 \rfloor_{p_{1,2}}
\lfloor 2j_{12}+2 \rfloor_{p_{1,2}}
\over
\lfloor p_{1,2} \rfloor \! !
}
\label{Xi}
\eeq
which we wrote, by anticipation, in a way that will be
 valid for continuous spins  since $p_{1,2}$ will remain an integer.
The symbol $\lfloor x \rfloor_n$ stands in general for
$\prod_{k=1}^{n}
\lfloor x+k-1 \rfloor$.
This  gives for half-integer spins
\beq
\Xi_{J_1J_2}^{J_{12}}
=
\sqrt{
\lfloor 2J_1 \rfloor \! !
\lfloor 2J_2 \rfloor \! !
\over
\lfloor 2J_{12}+1 \rfloor \! !
}
{1\over
\Delta(J_1,J_2,J_{12})
}.
\eeq
Using this, we compute
$$
\left\{ ^{\quad J_1}_{j+m_{12}}\,^{J_2}_j\right | \left .
\,^{J_{12}}_{j+m_1}\right\}
=
{
\Xi _{J_2j+m_{12}}^{j+m_1}\
\Xi_{{J_1}j+m_1}^{j}
\over
\Xi_{J_1J_2}^{J_{12}}\
\Xi_{J_{12}j+m_{12}}^{j}
}
e^{i\pi (J_1+J_2-2J_{12}-m_{12}-2j)}
\lfloor 2j+2m_1+1 \rfloor
$$
$$
(\Delta(J_2,j+m_1,j+m_{12})
\Delta(J_1,j,j+m_1))^2
$$
$$
\sum_{z}
{
e^{i\pi z}
\lfloor z+1 \rfloor \! !
\over
\lfloor z\!-\!J_1\!-\!J_2\!-\!J_{12} \rfloor \! !
\lfloor z\!-\!2j\!-\!J_1\!-\!m_1 \rfloor \! !
\lfloor z\!-\!2j\!-\!J_2\!-\!m_1\!-\!m_{12} \rfloor \! !
\lfloor z\!-\!2j\!-\!J_{12}\!-\!m_{12} \rfloor \! !
}
$$
\beq
{
1
\over
\lfloor J_1+J_2+m_{12}+2j-z \rfloor \! !
\lfloor J_1+J_{12}+m_1+m_{12}+2j-z \rfloor \! !
\lfloor J_2+J_{12}+m_1+2j-z \rfloor \! !
}
\! .
\label{sixj1b}
\eeq
The sum is for integer $z$ such that the factorials have positive
arguments\footnote{The terms $2j$ have been put on the other side of
the inequalities for later convenience.}, that is, for
\beq
\left \{
\begin{array}{cccc}
z-2j&\le& J_1+J_2+m_{12} &\hbox{(sup1)}\\
z-2j&\le& J_1+J_{12}+m_1+m_{12} &\hbox{(sup2)}\\
z-2j&\le& J_2+J_{12}+m_1 &\hbox{(sup3)}\\
z-2j&\ge& J_1+m_1&\hbox{(inf1)}\\
z-2j&\ge& J_{12}+m_{12} &\hbox{(inf2)}\\
z-2j&\ge& J_2+m_1+m_{12}&\hbox{(inf3)}\\
z&\ge& J_1+J_2+J_{12}& \hbox{(inf4)}\\
\end{array}
\right.
\label{bound1}.
\eeq
The TI3 conditions  can be seen to come from the conditions of
existence of a non-empty range for $z$:
the three upper bounds must individually be greater than
the four lower bounds,
which yields twelve inequalities that can be rearranged
in four TI3's.

This suggests the following generalization to non-half-integer $j$:
in this case, if $z$ were  kept integer,
the first six bounds would  not remain   integer
and should therefore be removed as they would  no longer be
given by (one over)  poles of gamma functions.
It seems much more sensible to choose  $z=2j+$ integer,
or equivalently to sum over $y\equiv z-2j=$ integer.
With this choice, it is the bound (inf4), on the contrary,
that is no longer integer (for $y$) and has to be removed.
We are left with the conditions
\beq
\left.
\begin{array}{ccc}
J_1+m_1 &\\
J_{12}+m_{12} &\\
J_2+m_1+m_2 & \\
\end{array}
\right\}
\le\ y=z-2j \ \le
\left\{
\begin{array}{c}
J_1+J_2+m_{12}\\
J_1+J_{12}+m_1+m_{12}\\
J_2+J_{12}+m_{12}\\
\end{array}
\right.
\label{bound2}.
\eeq

It is remarkable that
the three remaining upper and lower bounds
give nine inequalities
that can precisely be combined in the three TI2's  and one
TI3 of the   fusion-diagram\ref{FusTwoCond}.

The mechanism may be summarized as follows.
Choose
some particular TI3, TI2 or TI1 for the four vertices.
It follows that certain  (sum of)  spins
must be integers.
The generalization of the 6-j's
is obtained by only keeping
the integer bounds
among the seven of Eq.\ref{bound1}, with a ``good'' choice for $z$ (mod 1).
Then, the existence of a non-empty
range of summation turns out to be equivalent
to the positivity conditions of the
TI3, TI2 and TI1 type  initially chosen.
The same mechanism will work in the case of
the braiding with four TI2's
and braiding/fusing with four TI1's  (the second step of generalization).

Then, the remaining
factorials with non-integer arguments must be combined two by two,
in order to avoid q-deformed gamma functions,
in the sum as well as in the $\Delta$ functions.
It is already done in the $\Xi$ factors.
We perform the following substitutions
$$
\lfloor z+1 \rfloor \! !
/
\lfloor z-J_1-J_2-J_{12} \rfloor \! !
\ \to\
\lfloor z-J_1-J_2-J_{12}+1 \rfloor _{J_1+J_2+J_{12}+1}
$$
for two factorials in the sum,
and
$$
(\Delta(J,j,j+m))^2
\ \to\
\lfloor J+m \rfloor \! !
\lfloor J-m \rfloor \! !
/\lfloor 2j+m-J+1 \rfloor _{2J+1}
$$
for the others which only appear in the $\Delta$ functions.
This, together with the use of the new summation
variable $y\equiv z-2j$,
shows that these newly defined 6-j symbols are in fact
rational fractions in the variable
$e^{ihj}$ (or $j$ in the non-deformed case).
It can be checked that this generalized definition
gives back the standard one when applied to half-integer spins
such that the four TI3's  are satisfied, as
the lower bound of summation (inf4)
is restored by the poles of the corresponding
gamma function.

\vskip 3mm

This is only a (natural) definition, we have now to prove the polynomial
relations
on 6-j symbols. Let us  concentrate on
 the pentagonal relations  which are  the
basic identities  for fusion.
It is convenient to introduce  the following function of $e^{ihj}$,
where
the $J_i$'s and $m_i$'s  are fixed parameters, and $m_{ij}=m_i+m_j$,
$$
P(e^{ihj})\equiv
\sixjxi J_1,j+m_{12},J_2,j,J_{12},j+m_1
\sixjxi J_{12},j+m_{123},J_3,j,J_{123},j+m_{12}
$$
\beq
-\sum_{J_{23}}
\sixjxi J_2,j+m_{123},J_3,j+m_1,J_{23},j+m_{12}
\sixjxi J_1,j+m_{123},J_{23},j,J_{123},j+m_1
\sixjxi J_1,J_3,J_2,J_{123},J_{12},J_{23}
\label{penta1}.
\eeq
The range of summation over $J_{23}$ is fixed by the individual conditions
for the 6-j:
\beq
|J_2-J_3|,|J_1-J_{123}|,|m_{23}|\ \le\ J_{23}
\ \le\ J_2+J_3,J_1+J_{123}
\label{pbound1}.
\eeq
$P$ is a sum of rational fractions of $e^{ihj}$,
with  a range of summation which is independent from  $j$.
It is thus  a rational fraction of $e^{ihj}$.
If $2j$ is chosen to be an integer large enough,
all the TI3's  are fulfilled. Then,
the 6-j's reduce to  the standard ones,
the usual  pentagonal equation shows that $P(e^{ihj})=0$
(there is an extra upper bound
$2j+2m_1+m_2+m_3$ for $J_{23}$ in the
standard pentagonal equation,
but it is irrelevant for $j$ large enough).
Hence, $P$ is a rational
fraction
with an infinite number of
distinct zeros $e^{ihn/2}$ ($h/\pi$ is not rational).
This shows   that $P$ is identically $0$ and,  therefore,
the pentagonal relation holds  for any
$j$\footnote{
The relation always holds even though
for $j$ half-integer not large enough
some 6-j can have poles cancelled by zeros of others.}.

We repeat that the square roots of the $\Xi$ factors cause
no trouble as the $\Xi$ factors with one entry equal to $J_{23}$ cancel out
and the others can be factorized.

\vskip 0.5 cm

The braiding
\beq
\epsffile{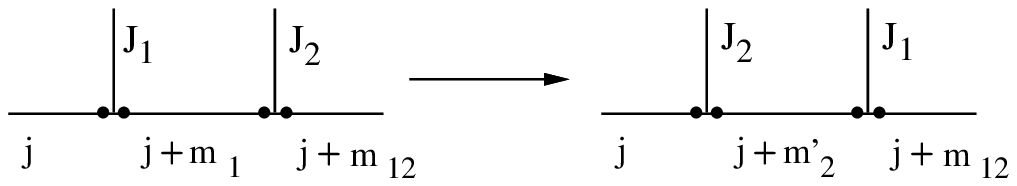}
\label{BrdTwoCond}
\eeq
leads to the definition of an other generalized 6-j coefficient.
As before,  the choice of a continuous $j$
leads  to choose $z-2j=y=$ integer.  This time
the upper bound (sup3) is the one  naturally released,
leaving only the four desired TI2.
We only write down the result:
$$
\sixjxi J_1,J_2,j+m_{12},j,j+m'_2,j+m_1 =
{
\Xi _{J_2j+m_{12}}^{j+m_1}
\Xi_{{J_1}j+m_1}^{j}
\over
\Xi_{J_1j+m_{12}}^{j+m'_2}
\Xi_{J_2j+m'_2}^{j}
}
e^{i\pi (J_1-J_2+m_{12}-2m'_2)}
\lfloor 2j+2m_1+1 \rfloor
\Delta(J_1,j,j+m_1)^2
$$
$$
(\Delta(J_2,j+m_1,j+m_{12}))^2
\sum_{y}
{
(-1)^y
\lfloor 2j+m_{12}+m_1+m'_2-y+1 \rfloor _{2y-m_{12}-m_1-m'_2}
\over
\lfloor y-J_1-m_1 \rfloor \! !
\lfloor y-J_1-m_{12}-m'_2 \rfloor \! !
}
\times
$$
\beq
{1\over
\lfloor y-J_2-m_1-m_{12} \rfloor \! !
\lfloor y-J_2-m'_2 \rfloor\! !
\lfloor J_1+J_2+m_{12}-y \rfloor \! !
\lfloor J_1+J_2+m_1+m'_2-y \rfloor \! !
}
\label{sixj2}
\eeq
for
\beq
\left.
\begin{array}{ccc}
J_1+m_1 &\\
J_1+m_{12}+m'_2 &\\
J_2+m'_2 & \\
J_2+m_{12}+m_1 & \\
\end{array}
\right\}
\le\ y\ \le
\left\{
\begin{array}{c}
J_1+J_2+m_{12}\\
J_1+J_2+m_1+m'_2\\
\end{array}
\right.
\label{bound3}.
\eeq

It is straightforward to prove the orthogonality
relation for generalized 6-j of this type,
using the same method as for the pentagonal relation.

\subsection{The second step of generalization}

We define now
(twice-generalized)
6-j coefficients for the fusion and braiding of operators
with only TI1 conditions
(figure \ref{OneCond}).
The fusion is then
\beq
\epsffile{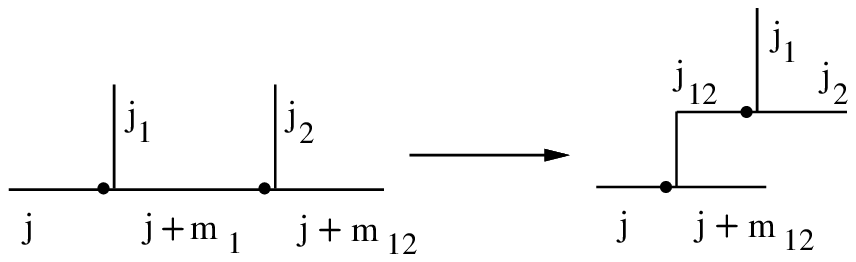}
\label{FusOneCond}.
\eeq
It now follows
from the general rules, that all spins are continuous\footnote{We shall
actually only use the case where they are rational. At this level of the
discussion this makes no difference (more about this below, however).},
and that the only conditions are
that   $j_i+m_i$, $i=1,2,12$ and $j_1+j_2-j_{12}$
be positive integers.

Now that   the principles of generalization have been written,
the twice-generalized 6-j will be easily deduced.
Among the bounds \ref{bound1},
only (inf1), (inf2), (sup1) and (sup2) are integers in the
case of four TI1's,
so that two more conditions (inf3) and (sup3) have to be released.
It is then easy to see that,  here again, the conditions of
existence of a non-empty range for $y$,
with these bounds (selected only according to their integer character),
precisely says that
$j_i+m_i$, $i=1,2,12$ and $j_1+j_2-j_{12}$  must be positive integers.

It is possible to
write the result without any gamma function
in the following rational form:
$$
\sixjxi j_1,j+m_{12},j_2,j,j_{12},j+m_1 =
{
\Xi _{j_2j+m_{12}}^{j+m_1}\
\Xi_{{j_1}j+m_1}^{j}
\over
\Xi_{j_1j_2}^{j_{12}}\
\Xi_{j_{12}j+m_{12}}^{j}
}
{
(-1)^{(j_1+m_1+j_2+m_2)}
\lfloor j_1+m_1 \rfloor \! !
\lfloor j_2+m_2 \rfloor \! !
\over
\lfloor 2j+m_1-j_1+1 \rfloor_{j_1+m_1}
\lfloor 2j+2m_1+2 \rfloor_{j_2+m_2}
}
\times
$$
$$
\sum_y
{
(-1)^y
\lfloor 2j+j_1+m_1+2 \rfloor _{y-(j_1+m_1)}
\lfloor 2j-j_1-j_2-j_{12}+y+1 \rfloor_{(j_1+m_1)+(j_{12}+m_{12})-y}
\over
\lfloor y-(j_{12}+m_{12} \rfloor \! !
\lfloor y-(j_1+m_1) \rfloor \! !
}
$$
\beq
{
\lfloor j_2+j_{12}+m_1-y+1 \rfloor_{y-(j_{12}+m_{12})}
\lfloor y-j_2-m_1-m_{12}+1 \rfloor_{(j_1+m_1)+(j_2+m_2)-y}
\over
\lfloor (j_1+m_1)+(j_2+m_2)-y \rfloor \! !
\lfloor (j_1+m_1)+(j_{12}+m_{12})-y \rfloor \! !
}
\label{sixj3}
\eeq
with
\beq
\left.
\begin{array}{ccc}
j_1+m_1 &\\
j_{12}+m_{12} &\\
\end{array}
\right\}
\le\ y\ \le
\left\{
\begin{array}{c}
(j_1+m_1)+(j_2+m_2)\\
(j_1+m_1)+(j_{12}+m_{12})\\
\end{array}
\right.
{}.
\eeq
This definition is similar up to normalization to
the definition of ref.\cite{AW}.
There, the generalized 6-j coefficient is expressed with a
$\,_4\! F_3$ hypergeometric function.
We give such an expression in section \ref{s4},
but here we prefer keeping an explicit sum in order to see more
clearly what are the bounds of summation,
as it is a key point of this generalization procedure.

This 6-j can be considered as a function of the three continuous
variables $e^{ihj}$, $e^{ihj_1}$, $e^{ihj_2}$,
and of three independent discrete variables chosen among
$j_1+m_1$, $j_2+m_2$, $j_{12}+m_{12}$
and $j_1+j_2-j_{12}$ (the latter  must  be positive integers
but are clearly not independent). For fixed values of these
discrete variables,
it is immediate to see that  the 6-j
 is
a rational fraction of the three continuous variables.

Going along a line similar to
the previous one, it is straightforward to check
that,  if $j_1$ and $j_2$ are chosen to be half-integers
large enough for the three TI2 and one TI3 of the first
step of generalization to be fulfilled,
the range of summation is reduced by the  gamma functions
so that  we get back the once-generalized  6-j of last subsection.

\vskip 3mm

Let us now prove the twice generalized pentagonal equation.
We define
$$
Q(e^{ihj_1},e^{ihj_2},e^{ihj_3})\equiv
\sixjxi j_1,j+m_{12},j_2,j,j_{12},j+m_1
\sixjxi j_{12},j+m_{123},j_3,j,j_{123},j+m_{12}
$$
\beq
-\sum_{j_{23}}
\sixjxi j_2,j+m_{123},j_3,j+m_1,j_{23},j+m_{12}
\sixjxi j_1,j+m_{123},j_{23},j,j_{123},j+m_1
\sixjxi j_1,j_3,j_2,j_{123},j_{12},j_{23}
\label{penta2}.
\eeq
We have to prove that $Q=0$.
Q is a function of the fixed discrete variables
$j_i+m_i$, $j_i+j_k-j_{ik}$
and of $e^{ihj}$,
that we keep fixed.

Here again
the range of summation for $j_{23}$ naturally
comes from the individual conditions
for the 6-j, and can be written:
$j_{23}=j_2+j_3-n$ for $n$ integer such that
\beq
0
\ \le\ n
\ \le\ (J_2+m_2)+(J_3+m_3),(j_1+j_2-j_{12})+(j_3+j_{12}-j_{123})
\label{pbound2}
\eeq
which shows that the range of summation only
depends on the fixed discrete variables.
Hence, $Q$  which is a sum on a fixed interval of rational fractions
of $e^{ihj_1}$, $e^{ihj_2}$ and $e^{ihj_3}$,
is a rational fraction as well.

Let us see that, for $j_1$, $j_2$, $j_3$
half-integers taken large enough,
$Q$ is equal to $P$ (of last subsection) and hence to zero.
With such $j_1$, $j_2$, $j_3$,
the twice-generalized 6-j's become once-generalized 6-j's,
as noted above.
So, we only have to prove that the range of summation over $j_{23}$ is
identical to the one (\ref{pbound1}) of $P$.
In terms of the better
suited variable $n\equiv j_2+j_3-j_{23}$,
the interval \ref{pbound1} is
\beq
0,j_2+j_3-j_1-j_{123}
\le n \le j_2+j_3-|j_2-j_3|,j_2+j_3-|j_2-j_3|,j_2+j_3-|m_{23}|
\label{pbound3}.
\eeq
One lower bound and two out of the six upper bounds
of \ref{pbound3}
(a bound with an absolute value being decomposed in two bounds)
are already verified thanks to Eq.\ref{pbound2}.
And the other bounds of \ref{pbound3}
are not relevant for $j_1$, $j_2$, $j_3$ large
enough:
noting that for fixed discrete variables ($j_i+m_i,...=$ constants)
$j_{123}$ is equal to $j_1+j_2+j_3+$constant, and $m_{23}$
to $-j_2-j_3+$constant,
we see that the extra bounds of \ref{pbound3}
are less restrictive than the bounds \ref{pbound2}.
For example $j_2+j_3-j_1-j_{123}=-2j_1+$constant
and is lower than the lower bound of \ref{pbound2}
for $j_1$ large enough.
It can easily be seen that for $j_1$, $j_2$, $j_3$ larger
than $(j_1+m_1)+(j_2+m_2)+(j_3+m_3)$ for example
(this is much larger than needed)
the bounds \ref{pbound1} are equivalent to \ref{pbound2}.
This proves that in this infinite number of cases $Q=P=0$.
Then, as $Q$ is a rational fraction, it is zero for any $j_1$, $j_2$, $j_3$,
which proves the pentagonal relation.

\vskip 5mm

The twice-generalized braiding 6-j
can be deduced from the once-generalized ones,
but it gives the same twice-generalized 6-j as fusion,
not surprisingly as both come from four TI1's  vertices
as shows the braiding of  TI1 operators:
\beq
\epsffile{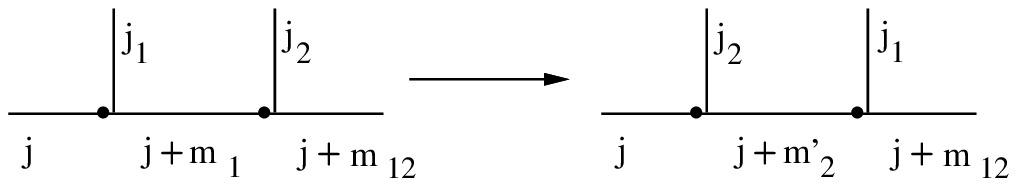}
\label{BrdOneCond}.
\eeq

The orthogonality relation for these (twice) generalized 6-j
can be easily proven using the same technique.
The other polynomial equations\cite{MS}
can as well be proven by this method of continuation
with the weights
$\Delta_j=-j(j+1)h/\pi$ for any $j$.
In the case of continuous spins,
orthogonality was the only polynomial equation proven in ref.\cite{AW},
and in the literature to our knowledge.

Our generalized
 polynomial equations have been numerically tested
using  random samples verifying only four TI1's
(in the q-deformed or classical cases).

This result has further applications:
it was proven in ref.\cite{CGR2} that letting some suitably
chosen spins go to infinity (with fixed differences),
the 6-j coefficients have Clebsch-Gordan coefficients or R-matrix
elements as limits.
The same is true for these generalized 6-j coefficients.
It defines Clebsch-Gordan coefficients and
R-matrix elements with continuous spins.
Performing this limit in polynomial equations
yields all the possible relations between 6-j's,
Clebsch-Gordan coefficients and R-matrix elements,
which can be found in ref.\cite{KR} in the case of
half-integer spins.
It allows to construct the basis of covariant operators ($\xi$)
with continuous spins as it is done in ref.\cite{GS3}.

\subsection{Other properties}

Our form \ref{sixj3} was derived for fusion and is
consequently  special  to this diagram.
We give the following symmetrized form which is  more
suitable to demonstrate polynomial equations
besides  the pentagonal one.
$$
\sixjxi j_1,j_3,j_2,j_{123},j_{12},j_{23} =
{
\Xi _{j_2j_3}^{j_{23}}\
\Xi_{{j_1}j_{23}}^{j_{123}}
\over
\Xi_{j_1j_2}^{j_{12}}\
\Xi_{j_{12}j_3}^{j_{123}}
}
{
(-1)^{p_{12,3}+p_{1,2}}
\lfloor p_{2,3} \rfloor \! !
\lfloor p_{1,23} \rfloor \! !
\lfloor 2j_{23}+1 \rfloor
\over
\lfloor j_{23}+j_{123}-j_1+1 \rfloor _{p_{1,23}+p_{2,3}+1}
}
$$
$$
\!\!\!\!\!\!\!\!\!\!\!\!
\!\!\!\!\!\!\!\!\!\!\!\!
\sum_{\qquad\qquad
p_{1,23},p_{12,3}\le y\le p_{12,3}+p_{1,23},p_{12,3}+p_{1,2}}
\!\!\!\!\!\!\!\!\!\!\!\!
\!\!\!\!\!\!\!\!\!\!\!\!
\!\!\!\!\!\!\!\!\!\!\!\!
\!\!\!\!\!\!\!\!
{
(-1)^y
\lfloor j_1+j_{23}+j_{123}+2 \rfloor_{y-p_{1,23}}
\lfloor y+2j_{123}-j_1-j_2-j_{12}+1 \rfloor_{p_{12,3}+p_{1,23}-y}
\over
\lfloor y-p_{1,23} \rfloor \! !
\lfloor y-p_{12,3} \rfloor \! !
}
$$
\beq
{
\lfloor y+2j_{123}-j_2-j_3-j_{23}+1 \rfloor_{p_{12,3}+p_{1,2}-y}
\lfloor j_2+j_{12}+j_{23}-j_{123}-y+1 \rfloor_{y-p_{12,3}}
\over
\lfloor p_{12,3}+p_{1,23}-y \rfloor \! !
\lfloor p_{12,3}+p_{1,2}-y \rfloor \! !
}
\label{sixj4}
\eeq
where the $p$'s are such that
\beq \left. \begin{array}{ll}
p_{1,2}\equiv j_1+j_2-j_{12}, &
p_{2,3}\equiv j_2+j_3-j_{23}, \nnn
p_{12,3}\equiv j_{12}+j_3-j_{123},&
p_{1,23}\equiv j_1+j_{23}-j_{123},
\end{array} \right \}  \rightarrow
\begin{array}{l}
p_{k,l}\in Z_+, \nnn
p_{1,2}+p_{12,3}=p_{1,23}+p_{2,3}.
\end{array}
\label{p}
\eeq

We note that we only have the residual symmetry
\beq
\left\{ ^{j_1}_{j_3}\,^{j_2}_{j_{123}}
\right. \left |^{j_{12}}_{j_{23}}\right\}
=
\left\{ ^{j_3}_{j_1}\,^{j_2}_{j_{123}}
\right. \left |^{j_{23}}_{j_{12}}\right\}
\label{symsixj}.
\eeq
The other symmetries are lost due to the particular choice
of the quantities $p_{k,l}$ to be positive integers.

We give the orthogonality relation:
\beq
\sum_{j_{123}-j_1\le j_{23}\le j_2+j_3}
\sixjxi j_1,j_3,j_2,j_{123},j_{12},j_{23}
\sixjxi j_3,j_1,j_2,j_{123},j_{23},j'_{12}
=
\delta_{j_{12},j'_{12}}
\label{orthog}
\eeq
with $j_{123}-j_3\le j_{12}\le j_1+j_2$
and $j_{123}-j_3\le j'_{12}\le j_1+j_2$.

\vskip 3mm

Then we study the properties of these 6-j with  some spins
shifted by $\alpha \pi/h$.
Since they are trigonometric functions of the spins multiplied by $h$,
and if the set of shifted spins is well chosen,
this only results in an overall sign for half-integer $\alpha$
--- or in nothing for integer $\alpha$.
These properties will be needed in next section.
We represent the shifts $s_i$ for every $j_i$ in an array recalling
the 6-j symbols for readability:
$
\left( ^{s_1}_{s_3}\,^{s_2}_{s_{123}}
\,^{s_{12}}_{s_{23}}\right)
$
{}.
First of all, if we want to have a simple relation
between the original and the shifted 6-j, we have
to keep the $p_{k,l}$ unchanged, as they control the range of summation.
This condition  gives four linear equations on the six shifts $s_i$,
leaving  three as independent. The latter may be chosen in different ways,
for instance,
$\left( ^{0}_{0}\,^{1}_{1}
\,^{1}_{1}\right)$,
$\left( ^{1}_{1}\,^{-1}_{1}
\,^{0}_{0}\right)
$, and
$\left( ^{1}_{-1}\,^{0}_{0}
\,^{1}_{-1}\right)
$. An equivalent choice is
$\left( ^{0}_{1}\,^{0}_{1}
\,^{0}_{1}\right),$
$\left( ^{1}_{0}\,^{0}_{1}
\,^{1}_{0}\right),$ and
$ \left( ^{-1}_{0}\,^{1}_{0}
\,^{0}_{1}\right)$.

These shifts however affect the linear combinations of spins that cannot
be written by means of $p_{k,l}$
and will only give a relation between both 6-j
if they are taken proportional
to $\alpha \pi/h$ with $2\alpha$ an integer,
thanks to $\lfloor x+n\pi/h \rfloor=(-1)^n \lfloor x \rfloor$
and the rules for square roots given above.
The actual calculation shows that the 6-j shifted by the above (array of)
shifts times $\pi/h$ times an integer are unchanged.
When the 6-j arguments are shifted by the above
shifts times $\pi/h$ times half an integer, they only get an extra sign,
at most.
We give it in three cases that will prove usefull in next section:
$$
\left\{\left( ^{j_1}_{j_3}\,^{j_2}_{j_{123}}
\right. \left |^{j_{12}}_{j_{23}} \right)
+\alpha{\pi\over h}
\left( ^{0}_{1}\,^{0}_{1}
\,^{0}_{1}\right)
\right\}
=
\left\{ ^{j_1}_{j_3}\,^{j_2}_{j_{123}}
\right. \left |^{j_{12}}_{j_{23}}\right\}
$$
$$
\left\{\left( ^{j_1}_{j_3}\,^{j_2}_{j_{123}}
\right. \left |^{j_{12}}_{j_{23}} \right)
+\alpha{\pi\over h}
\left( ^{1}_{0}\,^{0}_{1}
\,^{1}_{0}\right)
\right\}
=
\left\{ ^{j_1}_{j_3}\,^{j_2}_{j_{123}}
\right. \left |^{j_{12}}_{j_{23}}\right\}
$$
\beq
\left\{\left( ^{j_1}_{j_3}\,^{j_2}_{j_{123}}
\right. \left |^{j_{12}}_{j_{23}} \right)
+\alpha{\pi\over h}
\left( ^{0}_{0}\,^{1}_{1}
\,^{1}_{1}\right)
\right\}
=
(-1)^{2\alpha(j_{12}+j_{23}-j_2-j_{123})}
\left\{ ^{j_1}_{j_3}\,^{j_2}_{j_{123}}
\right. \left |^{j_{12}}_{j_{23}}\right\}
\label{shift}
\eeq
for $2\alpha\in Z$.

\section{$\!\!\!\!$ALGEBRA OF GENERALIZED OPERATORS}
\markboth{ 4. algebra of generalized operators}
{ 4. algebra of generalized operators}
\label{s3}

In this section,
we determine  the algebra of generalized vertex operators
using the quantum-group invariant operator basis,   in
two steps.
\subsection{The case  of $U_q(sl(2)). $}
The fusion and braiding matrices  of the standard operators,
 --- i.e. of operators with only   half-integers spins
verifying the full triangular inequalities (TI3), and
 represented on figure \ref{ThreeCond} ---
were  computed
recently in ref.\cite{CGR1} (see Eq.\ref{2.1}, with all hatted spins
set equal zero).
However, the operators considered by Gervais et al
from the beginning were not of this kind.
They were $V^{(J)}_m$ operators with $J+m\in Z_+$ and $J-m\in Z_+$,
but matrix elements with
arbitrary  zero-modes  were considered.
This situation corresponds to the case of only two conditions
 (TI2)
represented on figure \ref{TwoCond}.
Before ref.\cite{CGR1}, however,  one could only
elucidate the full algebra  in the basis
of the quantum group covariant
operators $\xi^{(J)}_m$ where the dependence in
the zero-mode disappears\cite{B,G1}.
Hence the first step of generalization of the algebra merely amounts to
finishing the work on well known operators,
proving that the braiding and fusing of these traditional
$V^{(J)}_m$ operators is essentially given by generalized 6-j
of the first step
(Eqs.\ref{sixj1b}, \ref{bound2} for fusion
and Eqs.\ref{sixj2}, \ref{bound3} for braiding).
Whereas it
was  useful in last section to introduce
the general method,
the first step of generalization  is   straightforward here,   and we shall
skip it for brevity sake.
The interested reader may deduce it by restriction\footnote{
keeping in mind the  caveats  given at the end of section 6,}  of the
second step to which we are going directly.
In the general case,   the operators are of a different kind, since
the quantum group spins  are no more half-integers.
The Coulomb-gas picture provides a convenient way to  build
the operators\cite{GS1}. One has
\beq
V^{(j)}_{m} \propto V^{(j)}_{-j} S^{j+m}
\label{scr}
\eeq
where $V^{(j)}_{-j}$ is the exponential of the  free
B\"acklund field, and $S$ is the screening operator.
This makes sense for arbitrary $j$, provided $j+m\in Z_+$,
 i.e. if the number of screening operators is a positive
integer.
 Remarkably, these
are our twice-generalized operators TI1, the ones with
only one  inequality
represented on figure \ref{OneCond}.
This correspondence is fully clarified in refs.\cite{GS1,GS3},
where the braiding of these  operators is computed.
Our next point is the fusion matrix    which  seems difficult to
compute, using the approach of refs.\cite{GS1,GS3} .
Of course, the fusion can be obtained from the braiding
by the three-leg symmetry following MS \cite{MS}.
We derive it nevertheless, for completeness and
in order to show how far one can go
by using the null-vector  equations.
As we shall see,
all the spins can be taken continuous except one (say $J_1$)
which has to be kept half-integer,
as there  must be at least  one
 degenerate conformal primary  (of the kind (1,$2J_1+1$))
so that the decoupling of the null vector
yields the differential equation which is the starting point.
We follow closely  the recursive proof of ref.\cite{CGR1},
generalizing it to non-half-integer spins,
this is why we will be sometimes sketchy
and refer the interested reader to the details given in ref.\cite{CGR1}.
We want to prove that these generalized operators
fit in Moore and Seiberg scheme\cite{MS}.
Thus their   fusion  must be of the form
$$
{\cal P}_j
V^{(j_1)}_{m_1}\,V^{(j_2)}_{m_2} =
{\cal P}_j
\sum_{j_{12}=-m_1-m_2}^{j_1+j_2}
F_{j+m_1,j_{12}}\!\!\left[
^{j_1}_j
\>^{\quad j_2}_{j+m_1+m_2 }
\right]\times
$$
\beq
\sum _{\{\nu_{12}\}} V ^{(j_{12},\{\nu_{12}\})} _{m_1+m_2}
<\!\varpi_{j_{12}},{\{\nu_{12}\}} \vert
V ^{(j_1)}_{j_2-j_{12}} \vert \varpi_{j_2} \! >,
\label{fus}
\eeq
where $j_i+m_i$ and $j_1+(j_2-j_{12})$ are positive integers
(these conditions determine  the summation range).

As said above,
we have to keep one spin half-integer. We choose
$J_1=1/2$ to begin the recursion.
It gives a degenarate family of the BPZ type (1,2) and
the differential  equation coming from the decoupling
of the null vector at level two allows us  to compute the
four-point function
$$<\varpi_{123} | V_{\pm 1/2}^{(1/2)}(z_1) V^{(j_2)}(z_2)
|\varpi_3> =z_2^{-\Delta_{j_2}-\Delta_{j_3}} \>
z_1^{-\Delta_{1/2}+\Delta_{j_{123}}}
\times
$$
\beq
 \left ( {z_2 \over
z_1}\right )^{\Delta_{j_{123}\pm 1/2} }  \left ( 1-{z_2 \over
z_1}\right )^{-hj_2 /\pi}\,_2 F_1(a_\pm ,b_\pm ;c_\pm ; {z_2\over z_1});
\label{f4p}
\eeq
$$
a_\pm ={1\over 2}+
{h\over 2\pi}\left [-\varpi_2  \mp (\varpi_{123}-\varpi_3 )\right
]
;\quad
b_\pm ={1\over 2}+ {h\over 2\pi}\left [-\varpi_2
 \mp (\varpi_{123}+\varpi_3 )
\right ]
;
$$
\beq
 c_\pm =1\mp{h\varpi_{123} \over \pi}
;\qquad
\varpi_i\equiv \varpi_{j_i}=\varpi_0+2j_i
\eeq
where $j_2,j_3,j_{123}$ can indeed be taken continuous.
The operator $V^{(j_2)}$ with continuous $j_2$
was not explicitely built in the Gervais et al approach yet
(hence the usefulness of the Coulomb-gas approach).
However, without this explicit construction,
if one assumes that such an operator exists,
the differential  equation allows to compute this four-point function.

The transformation properties of the
hypergeometric function $_2F_1$ in Eq.\ref{f4p}
allows us to show that the fusion of $V^{(1/2)}$ and $V^{(j)}$
are of the  MS form recalled above  with
$$
F_{j_{123}+\epsilon_1/2,j_2+\epsilon_2/2}\!
\left[^{1/2}_{j_{123}}\,^{j_2}_{j_3}\right]=
{
\Gamma(1-\epsilon_1\varpi_{123}h/\pi)
\over
\Gamma(1/2+(
-\epsilon_1\varpi_{123}+\epsilon_2\varpi_2+\varpi_3)h/2\pi)
}
$$
\beq
{
\Gamma(\epsilon_2\varpi_2h/\pi)
\over
\Gamma(1/2+(
-\epsilon_1\varpi_{123}+\epsilon_2\varpi_2-\varpi_3)h/2\pi)
}
\label{fus1/2}.
\eeq
Still following ref \cite{CGR1}, we try the ansatz
\beq
F_{{j_{23}},{j_{12}}}\!\!\left[^{j_1}_{j_{123}}
\,^{j_2}_{j_3}\right]
=
{g_{j_1j_2}^{j_{12}}\
g_{j_{12}j_3}^{j_{123}}
\over
g _{j_2j_3}^{j_{23}}\
g_{{j_1}j_{23}}^{j_{123}}
}
\left\{ ^{j_1}_{j_3}\,^{j_2}_{j_{123}}
\right. \left |^{j_{12}}_{j_{23}}\right\}.
\label{Fj}
\eeq
It is possible to see that the fusion coefficient
computed above can be cast  under the form:
\beq
F_{j_{123}+\epsilon_1/2,j_2+\epsilon_2/2}\!
\left[^{1/2}_{j_{123}}\,^{j_2}_{j_3}\right]=
{
g_{1/2\,j_2}^{j_2+\epsilon_2/2}\
g_{j_2+\epsilon_2/2\,j_3}^{j_{123}}
\over
g _{j_2j_3}^{j_{123}+\epsilon_1/2}\
g_{1/2\,j_{123}+\epsilon_1/2}^{j_{123}}
}
\left\{ ^{1/2}_{j_3}\,^{j_2}_{j_{123}}
\right. \left |^{j_2+\epsilon_2/2}_{j_{123}+\epsilon_1/2}\right\}
\label{F1/2}
\eeq
with (letting as usual $F(z)\equiv \Gamma(z)/\Gamma(1-z)$)
\beq
g_{j_1j_2}^{j_{12}}
= (g_0)^{j_1+j_2-j_{12}}
\prod_{k=1}^{j_1+j_2-j_{12}}
\sqrt{
F((\varpi_1-k)h/\pi)
F((\varpi_2-k)h/\pi)
F((-\varpi_{12}-k)h/\pi)
\over
F(1+kh/\pi)
}.
\label{gj}
\eeq
Then, we use the pentagonal equation for generalized
6-j symbols that was proven
in last section.
It would allow to make a recursion on the
integer quantities $j_i+j_k-j_{ik}$,
and compute the most general fusion coefficients.
However we are short of
  a starting point general enough for this recursion,
as our starting point (\ref{F1/2}),
the fusion of spin $1/2$ and a continuous spin $j_2$,
does not involve two continuous spins.
Hence, in order to be able to begin the recursion,
we have to only use pentagonal equations involving
fusion operations of the following type:
\beq
\epsffile{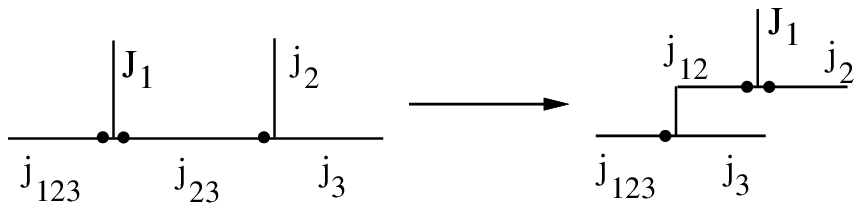}
\label{Fus1,5Cond}
\eeq
which can be seen as a restricted case of the twice-generalized one,
due to the half-integer character of $J_1$
(remember the dot convention of last section).
All the results of the second step of generalization apply,
with the extra condition that
$J_1+j_{123}-j_{23}$ and $J_1+j_{12}-j_2$ be positive.
Our starting point, the fusion coefficient \ref{F1/2},
corresponds to the fusion \ref{Fus1,5Cond} with $J_1=1/2$.
So, we make use of the following pentagonal relation:
$$
\epsffile{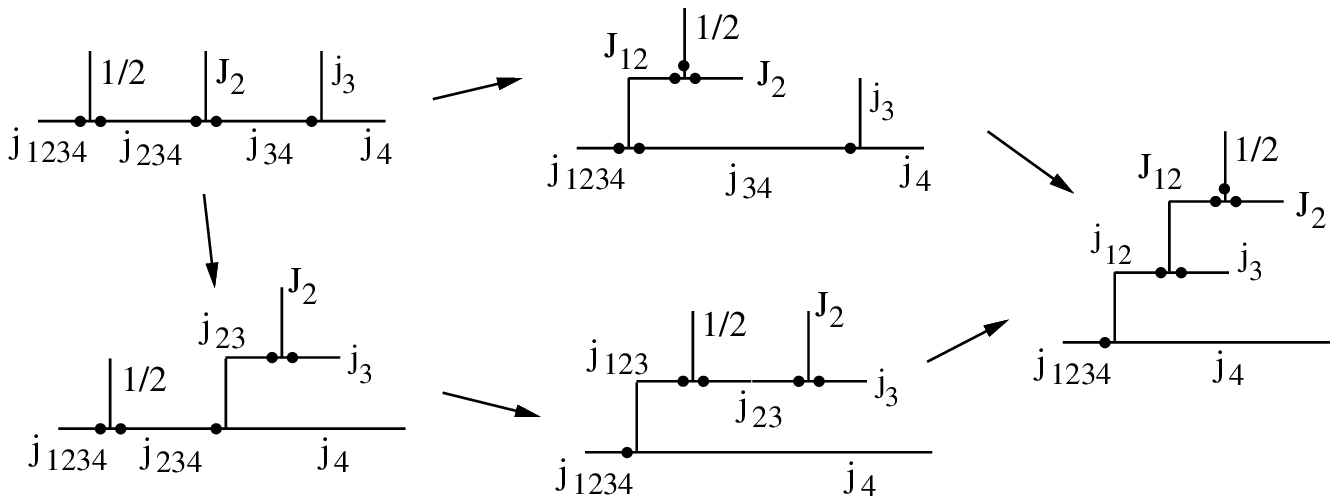}
$$
which only involves fusions of the type \ref{Fus1,5Cond}.
Beginning with $J_2=1/2$,
it is straightforward to show recursively with $J_2$ that
\beq
F_{{j_{34}},{j_{23}}}\!\!\left[^{J_2}_{j_{234}}
\,^{j_3}_{j_4}\right]
=
{g_{J_2j_3}^{j_{23}}\
g_{j_{23}j_4}^{j_{234}}
\over
g _{j_3j_4}^{j_{34}}\
g_{{J_2}j_{34}}^{j_{234}}
}
\left\{ ^{J_2}_{j_4}\,^{j_3}_{j_{234}}
\right. \left |^{j_{23}}_{j_{34}}\right\}
\label{FJj}
\eeq
with the restrictions (in comparison with the most general case)
that $J_2$ be half-integer and that $J_2+j_{234}-j_{34}$ and $J_2+j_{23}-j_3$
be positive.
The final step to the ansatz \ref{Fj} is then a conjecture.
It seems however very reasonable as it is only the result of the
symmetrization of Eq.\ref{FJj} to 6-j symbols that satisfy the
required polynomial equations.
Moreover it is confirmed   by the
result already proven for braiding in ref.\cite{GS1}.

\subsection{The complete algebra}

We have done   half of the work only,  since
there are two different screening charges, or
equivalently two deformation parameters
such that $h/\pi=(\alpha_-)^2/2$
and $\hhat/\pi=(\alpha_+)^2/2$.
In the half-integer case they give the BPZ degenerate families
$(1,2J+1)$ and $(2\Jhat+1,1$) respectively,
and all the families $(2J+1,2\Jhat+1)$ by fusion.
This is the double $U_q(sl(2))\odot U_\qhat (sl(2))$ structure
recalled  in the introduction.
The half-integer spin operators
of the full algebra
$V^{(J,\Jhat)}_{m,\mhat}$
were described by the four dicrete quantum numbers
$J,\Jhat,m,\mhat$.
The question is to know what  will be  the good quantum numbers in
the continuous spin case.
Let us first stick to the half-integer spin case for a short while.
The Coulomb gas realization of ref.\cite{GS1,GS3}
with both screening charges is given by
\beq
V^{(J,\Jhat)}_{m,\mhat} \propto V^{(J)}_{-J}
\Vhat^{(\Jhat)}_{-\Jhat} S^{J+m} \Shat^{\Jhat+\mhat}
\label{scr2}.
\eeq
Since $V^{(J)}_{-J}$, and $\Vhat^{(\Jhat)}_{-\Jhat}$ are exponentials of
the same free field, one has
\beq
V^{(J)}_{-J} \Vhat^{(\Jhat)}_{-\Jhat} \propto
V^{(J+\Jhat \pi/h)}_{-J-\Jhat\pi/h },
\label{scr3}
\eeq
so that $V^{(J,\Jhat)}_{m,\mhat}$ may be regarded as  a function of
the combination $\Je\equiv J+\Jhat\pi/h$,
that we call effective spin,  and not of $J$ and $\Jhat$ separately.
Concerning the states, a similar phenomenon occurs.
For  a state associated with   half-integer spins
$J$ and $\Jhat$, the corresponding zero-mode is
$\varpi=\varpi_{J,\Jhat}\equiv \varpi_0+2J+2\Jhat\pi/h$.
It is only a function of  $\Je$
and  we have $\varpi_{J,\Jhat}=\varpi_0+2\Je$, also
denoted $\varpi_{\Je}$.
One may verify that
 the fusion and braiding matrices may be written  in
terms of these  effective spins (more about this below).
Of course, if $h$ is irrational, using $J$, $\Jhat$ or $\Je$
is immaterial.
In this  half-integer spin case,
there is a remarkable fact, which is related:
by using the properties of the gamma functions under integer shifts
and of the sinus functions under shifts by ($\pi\times$integer),
one can show that any
  operators of the $U_q(sl(2))$ family
 has a trivial braiding and fusion matrices  ---
simple sign factors ---  with any operator
of the $U_{\qhat}(sl(2))$ type
(see refs.\cite{G1,G3,CGR1} and references therein).
This is the meaning of the $\odot$ in
$U_q(sl(2))\odot U_\qhat (sl(2))$, it is  a sort of graded
 tensorial product. These  sign factors  simply come out when one
returns from the effective spins to the half-integer ones.

For continuous case, the operators are $V^{(\Je)}_{-\Je}
S^{J+m} \Shat^{\Jhat+\mhat}$. They are specified by $\Je$ and the two
screening numbers $J+m$, and $\Jhat+\mhat$ that are positive integers.
The hatted and unhatted quantum numbers $J$ and $\Jhat$
can no longer be separated,
and  the effective spins $\Je$ turn out to be the only
possible good variables. As a result the relative fusion and braiding
of the two families with a single screening charge,  become non trivial,
as shown in ref.\cite{GS3}, and as we shall see.
So, we review what the TI3 and TI1 vertices become
in the case of the full algebra,
in terms of effective spins.
Treating the three legs symmetrically
(instead of using the shift quantities
$m$),
the TI3 conditions  for the full algebra
may  be written as\footnote{
$Z_++(\pi/h)Z_+\equiv\{a+b\pi/h\ ; \ a,b\hbox{ positive integers}\}$}
\beq
\epsffile{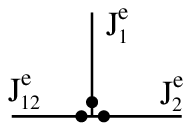}
\to
\left\{
\begin{array}{ccc}
\Je_1+\Je_2-\Je_{12}  \in Z_++(\pi/h)Z_+\\
\Je_{12}+\Je_2-\Je_1  \in Z_++(\pi/h)Z_+\\
\Je_1+\Je_{12}-\Je_2  \in Z_++(\pi/h)Z_+\\
\end{array}
\right\}
\ \Rightarrow\
2\Je_1,2\Je_2,2\Je_{12} \in Z_++{\pi\over h}Z_+
\label{ThreeCondEff}.
\eeq
These conditions are clearly equivalent to the
conditions \ref{ThreeCond} applied to unhatted  and hatted spins.
Now that the good variables have been
chosen, the generalization to continuous spins works
for $U_q(sl(2))\odot U_\qhat (sl(2))$
  just like in the case of the simple algebra $U_q(sl(2))$.
The generalized vertex is
\beq
\epsffile{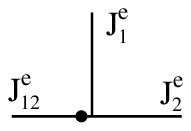}
\ \to\
\Je_1+\Je_2-\Je_{12} \in Z_++{\pi\over h}Z_+
\label{OneCondEff}.
\eeq
More generally, we extend the notation of Eq.\ref{p}
to the full algebra, and for every vertex $\,^{\Je_k}_{\Je_{kl}\Je_l}$,
we introduce the quantities
$p_{k,l}$ and $\phat_{k,l}$ defined by
$j_k+j_l-j_{kl}\equiv p_{k,l}+\phat_{k,l}(\pi/h)$
which are constrained to be positive integers\footnote{
This notation assumes a specific choice of labels for
the six $j$ parameters,
but it has the advantage of simplicity.}.
The vertex \ref{OneCondEff} represents the free field raised
at a continuous power $\Je_1$
screened by integer numbers $p_{1,2}$ and $\phat_{1,2}$
of the two different screening operators.

It leads us to the following ansatz:
\beq
F_{{\Je_{23}},{\Je_{12}}}\!\!\left[^{\Je_1}_{\Je_{123}}
\,^{\Je_2}_{\Je_3}\right]
=
{g_{\Je_1\Je_2}^{\Je_{12}}\
g_{\Je_{12}\Je_3}^{\Je_{123}}
\over
g _{\Je_2\Je_3}^{\Je_{23}}\
g_{{\Je_1}\Je_{23}}^{\Je_{123}}
}
\left\{\left\{ ^{\Je_1}_{\Je_3}\,^{\Je_2}_{\Je_{123}}
\left. \right |^{\Je_{12}}_{\Je_{23}}\right\}\right\}
\gaghat\gaghat\, ^{\Jehat_1}_{\Jehat_3}\,^{\Jehat_2}_{\Jehat_{123}}
\bverthat\bigr. \, ^{\Jehat_{12}}_{\Jehat_{23}}\gadhat\gadhat,
\label{Fjjhat}
\eeq
whose precise definition follows next.
The $\Jehat_i$'s are rescaled from the $\Je_i$'s by
$\Jehat_i=\Je_i h/\pi$ as usual.
The fusion matrix $F$ and the $g$ coefficients, although
noted as before, are a generalization of the previous ones with the condition
Eq.\ref{OneCondEff}
for the full algebra vertices.
These $g$'s have to be computed.
The 6-j symbols in Eq.\ref{Fjjhat} are noted with double braces,
which means that they are (slightly) different from
the ones of last section.
This is of course necessary as the $\Je_i$'s verify the
conditions \ref{OneCondEff} of the full algebra type
whereas the 6-j-symbol arguments must verify standard
conditions of the type \ref{p}.
We shall call them effective 6-j coefficients.
One may verify that we will  satisfy these standard conditions if we shift
the effective $J$'s  by suitable
linear combinations  either of  $p_{k,l}$'s or of
the $\phat_{k,l}$'s.
In fact, for 6-j coefficients computed
with one deformation parameter --- i.e. one of the last two factors
 on  the r.h.s. of Eq.\ref{Fjjhat}, say the first one ---
it is convenient  to shift the  $\Je_i$'s  by linear combinations
of the $\phat_{k,l}$'s, so that
conditions \ref{ThreeCondEff}
reduces to  conditions Eqs.\ref{p}  that only invole the  $p_{k,l}$'s.
This leads to the possible definition
$$
\left\{
\left\{ ^{\Je_1}_{\Je_3}\,^{\Je_2}_{\Je_{123}}
\right. \left |^{\Je_{12}}_{\Je_{23}}\right\}\right\}
\equiv
\left\{ ^{\Je_1}_{\Je_3}\,^{\Je_2}_{\Je_{123}+(\phat_{2,3}+\phat_{1,23})\pi/h}
\right. \left |^{\Je_{12}+\phat_{1,2}\pi/h}_
{\Je_{23}+\phat_{2,3}\pi/h}\right\}
$$
\beq
=
\left\{ ^{J_1+\Jhat_1\pi/h}_{J_3+\Jhat_3\pi/h}
\,^{\qquad J_2+\Jhat_2\pi/h}_{J_{123}+(\Jhat_1+\Jhat_2+\Jhat_3)\pi/h}
\right. \left |^{J_{12}+(\Jhat_1+\Jhat_2)\pi/h}
_{J_{23}+(\Jhat_2+\Jhat_3)\pi/h}\right\}
\label{sixjp}
\eeq
since one can easily see that the spins of the
6-j symbol of the r.h.s. satisfy the
conditions \ref{p} of the half-algebra.
A potential   trouble is that there are many possible choices,
none of them seemingly  better than the other.
The solution of the puzzle is that all the choices give the same value:
$$
\left\{\left\{ ^{\Je_1}_{\Je_3}\,^{\Je_2}_{\Je_{123}}
\right. \left |^{\Je_{12}}_{\Je_{23}}\right\}\right\}
=
\left\{ ^{\qquad
\Je_1}_{\Je_3-(\phat_{2,3}+\phat_{1,23})\pi/h}\,^{\Je_2}_{\Je_{123}}
\right. \left
|^{\Je_{12}+\phat_{1,2}\pi/h}_{\Je_{23}-\phat_{1,23}\pi/h}\right\}
$$
\beq
=
\left\{ ^{\Je_1-\phat_{1,23}\pi/h}_{\Je_3-\phat_{12,3}\pi/h}
\,^{\Je_2+(\phat_{1,23}-\phat_{1,2})\pi/h}_{\qquad \Je_{123}}
\right. \left |^{\Je_{12}}_{\Je_{23}}\right\}
=...,
\label{sixjp2}
\eeq
where we wrote various possibilities.
These equalities  are   easy consequences  of the properties proven at the
end of section \ref{s2}.
This result
is not surprising as the differences between the different choices
are precisely shifts of the spins  by (integer$\times \pi/h$),
 such that the $p_{k,l}$ are unchanged, which are
precisely the kind of shifts that were proven not to change the value
of the 6-j coefficients in section 3.3.
For instance, we compare the choice of Eq.\ref{sixjp}
and the first one of Eq.\ref{sixjp2},
(using the notations of subsection \ref{s2}.3 for the addition of sixtuples):
$$
\left( ^{\Je_1}_{\Je_3}\,^{\qquad
\Je_2}_{\Je_{123}+(\phat_{2,3}+\phat_{1,23})\pi/h}
\!\right. \left
|^{\Je_{12}+\phat_{1,2}\pi/h}_{\Je_{23}+\phat_{2,3}\pi/h}\right)
$$
$$
=
\left( ^{\qquad
\Je_1}_{\Je_3-(\phat_{2,3}+\phat_{1,23})\pi/h}\,^{\Je_2}_{\Je_{123}}
\!\right. \left
|^{\Je_{12}+\phat_{1,2}\pi/h}_{\Je_{23}-\phat_{1,23}\pi/h}\right)
+
(\phat_{2,3}+\phat_{1,23})
{\pi\over h}
\left( ^{0}_{1}\,^{0}_{1}
\,^{0}_{1}\right)
$$
which proves the equality of the corresponding 6-j coefficients
thanks to Eqs.\ref{shift}.

This invariance of the 6-j coefficients under some particular shifts
of their arguments allows as well to show that the ansatz Eq.\ref{Fjjhat}
verifies the pentagonal equation.
In these     equations,
the $g$ coefficients cancel out and the sum to be proven
factorises in a sum over $p_{2,3}$ of effective 6-j's
and a sum over $\phat_{2,3}$ of hatted effective 6-j's.
The difficulty is that when replacing the effective 6-j's by one
of their definitions in terms of (normal) 6-j's,
all the 6-j's do not have the same $\Je_i$'s,
they are shifted by some $\phat_{k,l}\pi/h$,
and the pentagonal equation \ref{penta2} cannot be
applied directly.
It is however easy to see that thanks to some shifts
of the previous kind,
it reduces to the pentagonal equation \ref{penta2}.

So, now that this ansatz has been proven to be consistently defined,
let us demonstrate it.
Let us get rid of the value of the $g$ coefficients first.
The fusion of a spin-one-half operator
  and of a generalized one  belonging to the
 the $U_q(sl(2))\odot U_\qhat (sl(2))$
algebra,
 involves two $g$ coefficients of the most general type.
There is no difficulty in computing the corresponding four-point
functions and to obtain this fusion coefficient
by
transformation of the gamma function.
It yields consistent necessary conditions for $g$ which
turns out to be the most natural generalization\footnote{The sign
$(-1)^{p\phat}$ was added in comparison with ref.\cite{CGR1}
(more about signs below).}
of the integer spin case of ref.\cite{CGR1}:
\beq
g_{\Je_1,\Je_2}^{\Je_{12}}= (-1)^{p\phat}
(i/2)^{p+\phat}{H_{p\phat}(\varpi_{\Je_1})H_{p\phat}(\varpi_{\Je_2})
H_{p\phat}(-\varpi_{\Je_{12}})
\over{H_{p\phat}(\varpi_{p/2,\phat /2})}}
\label{ggen}
\eeq
with
\beq
H_{p\phat}(\varpi) =
{
\prod_{r=1}^p \sqrt{F(\varpihat -r h/\pi )}
\prod_{\rhat =1}^{\phat}
\sqrt{F(\varpi - \rhat \pi /h )}
\over
\prod_{r=1}^p \prod_{\rhat =1 }^{\phat }
\left ( \varpi \sqrt{h/\pi}
-r \sqrt{h/\pi} -\rhat \sqrt{\pi /h}\right ) }
\label{H}.
\eeq
It can as well be written by absorbing the denominator factors
$$
H_{p\phat}(\varpi) =
\prod_{i=1}^{n-1}
\left\{
F\left[\varpi-N_i-(\Nhat_i+{1+\epsilonhat_i\over2}){\pi\over h}\right]
\left({-\pi\over h}\right)^{N_i}
\right\}^{\epsilonhat_i/2}
$$
\beq
\left\{
F\left[\varpihat-\Nhat_i-(N_i+{1+\epsilon_i\over2}){h\over \pi}\right]
\left({-h\over\pi}\right)^{\Nhat_i}
\right\}^{\epsilon_i/2}
\label{H2}
\eeq
where the ${(N_i,\Nhat_i),i=1...n}$ describe an arbitrary planar path
going from $(0,0)=(N_1,\Nhat_1)$ to $(p,\phat)=(N_n,\Nhat_n)$.
The allowed elementary steps $(\epsilon_i,\epsilonhat_i)\equiv
(N_{i+1}-N_i,\Nhat_{i+1}-\Nhat_i)$
are $(0,\pm 1)$ and $(\pm 1,0)$.

Now the work is almost finished.
Applying\footnote{
For brevity sake, we use in this paragraph the following short notation:
$V^{(J_i)}$ (resp. $\Vhat^{(\Jhat_i)}$) is a half-integer spin operator
built with deformation parameter $h$ (resp. $\hhat$),
$V^{(j_i)}$ (resp. $\Vhat^{(\jhat_i)}$) is a continuous
spin operator of the half-algebra with deformation parameter $h$ (resp.
$\hhat$),
hence verifying $j_i+j_{right}-j_{left}\in Z_+$,
and $V^{(\Je_i)}$ is a continuous
spin operator of the full algebra,
hence verifying $\Je_i+\Je_{right}-\Je_{left}\in Z_++(\pi/h)Z_+$.
}
the pentagonal equation (i.e. fusion associativity) to
$V^{(1/2)}V^{(J_2)}
V^{(\Je_3)}$,
the fusion coefficient of $V^{(1/2)}V^{(\Je)}$
allows to begin the recursion and prove that the one of
$V^{(J_1)}V^{(\Je_2)}$ verifies the ansatz Eq.\ref{Fjjhat}.
We conjecture that such is the case for
$V^{(j_1)}V^{(\Je_2)}$ as well.
We do the same for
$\Vhat^{(\jhat_1)}V^{(\Je_2)}$.
And eventually,
the pentagonal equation applied to
$V^{(j_1)}\Vhat^{(\jhat_2)}V^{(\Je_3)}$
proves the ansatz:
the fusion of $V^{(\Je_1)}V^{(\Je_2)}$.

\vskip 3mm

This result looks quite different from
the one for half-integer spins of ref.\cite{CGR1}.
There, the 6-j coefficients had entries $J_i$
and $\Jhat_i$ for the hatted 6-j.
The only mixing between hatted and unhatted quantities
was lying (in the $g$ factors and) in
an extra sign $(-1)^{\fusV 1,2,123,23,3,12 }$ with:
\beq
\fusV 1,2,123,23,3,12
=2\Jhat_2(J_{12}+J_{23}-J_2-J_{123})
+2J_2(\Jhat_{12}+\Jhat_{23}-\Jhat_2-\Jhat_{123})
\label{sign}.
\eeq
A parenthetical remark is in order at this point. The expression just
given does not agree with the one of ref.\cite{CGR1}. The latter was
derived  with a treatment of signs of square roots that is not completely
consistent, so that it is not really correct. This does not matter
for the polynomial equations, since the difference may be
re-absorbed by a change of coupling constants
$$
g_{\Jgen1 \Jgen2 }^{\Jgen{12} } \to
(-1)^{2p_{12} \Jhat_{12}+2\phat_{12}J_{12}}
g_{\Jgen1 \Jgen2 }^{\Jgen{12} },
$$
but Eq.\ref{sign} is the one which is completely consistent.

We have to check that our result for continuous spins
  reduces to this (corrected)
 old one when
applied to spins $\Je_i=J_i+\Jhat_i\pi/h$.
The $g$ coefficients are trivially the restriction of ours.
The non-trivial work is to prove
$$
\left\{
\left\{ ^{J_1+\Jhat_1\pi/h}_{J_3+\Jhat_3\pi/h}
\,^{J_2+\Jhat_2\pi/h}_{J_{123}+\Jhat_{123}\pi/h}
\right. \left |^{J_{12}+\Jhat_{12}\pi/h}
_{J_{23}+\Jhat_{23}\pi/h}\right\}\right\}
\gaghat\gaghat\, ^{\Jhat_1+J_1\pi/\hhat}_{\Jhat_3+J_3\pi/\hhat}
\,^{\Jhat_2+J_2\pi/\hhat}_{\Jhat_{123}+J_{123}\pi/\hhat}
\bigr. \bverthat\, ^{\Jhat_{12}+J_{12}\pi/\hhat}
_{\Jhat_{23}+J_{23}\pi/\hhat}\gadhat\gadhat
=
$$
\beq
(-1)^{\fusV 1,2,123,23,3,12 }
\left\{ ^{J_1}_{J_3}\,^{J_2}_{J_{123}}
\right. \left |^{J_{12}}_{J_{23}}\right\}
\gaghat\, ^{\Jhat_1}_{\Jhat_3}\,^{\Jhat_2}_{\Jhat_{123}}
\bigr. \bverthat\, ^{\Jhat_{12}}_{\Jhat_{23}}\gadhat .
\label{comparison}
\eeq
By applying the definition \ref{sixjp}, for example,
to
the first effective 6-j of Eq.\ref{comparison},
and noticing that
$$
\left( ^{J_1+\Jhat_1\pi/h}_{J_3+\Jhat_3\pi/h}
\,^{\ J_2+\Jhat_2\pi/h}_{J_{123}+(\Jhat_1+\Jhat_2+\Jhat_3)\pi/h}
\right. \left
|^{J_{12}+(\Jhat_1+\Jhat_2)\pi/h}_{J_{23}+(\Jhat_2+\Jhat_3)\pi/h}\right)
$$
$$
=
\left( ^{J_1}_{J_3}\,^{J_2}_{J_{123}}
\right. \left |^{J_{12}}_{J_{23}}\right)
+
\Jhat_1{\pi\over h}
\left( ^{1}_{0}\,^{0}_{1}
\,^{1}_{0}\right)
+
\Jhat_2{\pi\over h}
\left( ^{0}_{0}\,^{1}_{1}
\,^{1}_{1}\right)
+
\Jhat_3{\pi\over h}
\left( ^{0}_{1}\,^{0}_{1}
\,^{0}_{1}\right),
$$
one can see that this 6-j ($J_i+\Jhat_i\pi/h$) reduces to the
half-integer spin 6-j ($J_i$) up to a sign thanks to Eqs.\ref{shift}.
Doing the same  for the hatted 6-j and collecting all the signs,
one easily completes the proof.

\vskip 3mm

For completeness, we write the full fusion-equation
$$
{\cal P}_{\Je_{123} }
V^{(\Je_1)} {\cal P}_{\Je_{23}} V^{(\Je_2)}
{\cal P}_{\Je_3 }
=\sum _{\Je_{12}}
{g_{\Je_1\Je_2}^{\Je_{12}}\
g_{\Je_{12}\Je_3}^{\Je_{123}}
\over
g_{\Je_2\Je_3}^{\Je_{23}}\
g_{{\Je_1}\Je_{23}}^{\Je_{123}}
}
\left\{\left\{ ^{\Je_1}_{\Je_3}\,^{\Je_2}_{\Je_{123}}
\right. \left |^{\Je_{12}}_{\Je_{23}}\right\}\right\}
\gaghat\gaghat\, ^{\Jehat_1}_{\Jehat_3}\,^{\Jehat_2}_{\Jehat_{123}}
\bigr. \bverthat\, ^{\Jehat_{12}}_{\Jehat_{23}}\gadhat\gadhat
$$
\beq
\sum _{\{\nu_{12}\}}
{\cal P}_{\Je_{123} }
V ^{(\Je_{12},\{\nu_{12}\})}
{\cal P}_{\Je_3 }
<\!\varpi_{\Je_{12}},{\{\nu_{12}\}} \vert
V ^{(\Je_1)} \vert \varpi_{\Je_2} \! >.
\label{fusion}
\eeq
The sum over $\Je_{12}$ is a sum over all the values of $\Je_{12}$ allowed
by the four full-algebra vertex-conditions of the type Eq.\ref{OneCondEff}.
It can be viewed as a double sum on $p_{1,2}$ and $\phat_{1,2}$.
The value of the $g$ coefficients for the full
algebra is given in Eq.\ref{ggen}.
The definition of the effective 6-j coefficients (double brace)
is given in Eqs.\ref{sixjp}, \ref{sixjp2}.

We give the braiding equation as well.
It was computed in ref.\cite{GS1} for one half of the algebra
and in ref.\cite{GS3}
for the full algebra.
It can be deduced from the fusion
by the three leg symmetry of the vertices\cite{MS,CGR1}:
\beq
<\varpi_{12} |  V^{(\Je_1)}
|\varpi_2>
= e^{i\pi(\Delta(\Je_1)+\Delta(\Je_2)-\Delta(\Je_{12}))}
<\varpi_{\Je_{12}} |  V^{(\Je_2)} |\varpi_1>.
\label{sym}
\eeq
This  gives
$$
{\cal P}_{\Je_{123} }
V^{(\Je_1)} {\cal P}_{\Je_{23}}
V^{(\Je_2)}
{\cal P}_{\Je_3 }
=
\sum_{\Je_{13}} e^{\pm i\pi (\Delta_{\Je_{123}}+\Delta_{\Je_3}
-\Delta_{\Je_{23}}-\Delta_{\Je_{13}})} \times
$$
\beq
{g_{\Je_1 \Je_3}^{\Je_{13}} g_{\Je_{13} \Je_2}^{\Je_{123}} \over
g_{\Je_2 \Je_3}^{\Je_{23}} g_{\Je_1 \Je_{23}}^{\Je_{123}}}
\left\{\left\{
^{\Je_1}_{\Je_2}\,^{\Je_3}_{\Je_{123}}
\right. \left |^{\Je_{13}}_{\Je_{23}}\right\}\right\}
\gaghat\gaghat
\,^{\Jehat_1}_{\Jehat_2}\,^{\Jehat_3}_{\Jehat_{123}}
\bigr. \bverthat\, ^{\Jehat_{13}}_{\Jehat_{23}}\gadhat\gadhat
{\cal P}_{\Je_{123} }
V ^{(\Je_2)} {\cal P}_{\Je_{13}}
V^{(\Je_1)}
{\cal P}_{\Je_3 }
\label{braiding}
\eeq
where again the sum over $\Je_{13}$ is to be understood as a double sum.

These new expressions have the virtue that, besides
the extension to continuous spins,
they give   expressions  of the fusing and braiding coefficients
which are analytic in the spins.
It did not seem easy to derive formulae  of this type
from the expression for half-integer spins
due to the sign $(-1)^{f_V}$. The analytic expression
for the $g$ coefficients is of great importance as well,
 as we shall see in section 7:  it
 directly leads  to the three-point function.

\section{THE SYMMETRY $j\to -j-1$}
\markboth{ 5. The symmetry $j\to -j-1$}
{ 5. The symmetry $j\to -j-1$}
\label{s4}

In this section, we study the behaviour
of the 6-j coefficients under the  transformation
$j_i\to -j_i-1$ applied to all the spins.
We prove that there exists a suitable continuation
for which the 6-j coefficients are unchanged.
This part  essentially follows the line of  ref.\cite{G3}, but  deals
with higher quantum-group symbols.

The operators with transformed spins $-j_i-1$
are needed in next section to build real-positive-weight operators
in the strong coupling regime \cite{G3}.
They are  needed in the weak coupling regime \cite{G5}, as well.
The basic feature in this transformation is
that the screening numbers $p_{k,l}$ of the four vertex operators involved
are transformed into $-p_{k,l}-1$, and so,
this symmetry defines  the fusion coefficient of vertex operators
with negative screening numbers.

The transformation $j_i\to -j_i-1$ will be performed
on an expression of the 6-j coefficients in terms of
$_4F_3$ hypergeometric function.
This is why
we shall first prove the following useful identity on
q-deformed $_4F_3$ hypergeometric functions, following a method
explained for instance in refs.\cite{A,S}:
\beq
\,_4F_3\left(
^{x,y,z,-N;}
_{a,b,c;}
1
\right)
=
{
\lfloor b-z \rfloor_N
\lfloor c-z \rfloor_N
\over
\lfloor b \rfloor_N
\lfloor c \rfloor_N
}
\,_4F_3\left(
^{a-x,a-y,z,-N;}
_{a,z-N-b+1,z-N-c+1;}
1
\right),
\label{F43}
\eeq
with the conditions that
$N$ be a positive integer (i.e. that the series be terminating) and
that $x+y+z-N+1=a+b+c$ (i.e. that the series be Saalschutzian).
Our conventions for the hypergeometric functions
are the following\footnote{They are borrowed from ref.\cite{G3}. The
appendix B of this reference shows the connection with the
usual mathematical notation. For the case considered, the two agree.}
\beq
\,_4F_3\left(
^{\alpha,\beta,\gamma,\delta;}
_{a,b,c;}
z
\right)
\equiv
\sum_{\nu=0}^{\infty}
{
\lfloor \alpha \rfloor_\nu
\lfloor \beta \rfloor_\nu
\lfloor \gamma \rfloor_\nu
\lfloor \delta \rfloor_\nu
\over
\lfloor a \rfloor_\nu
\lfloor b \rfloor_\nu
\lfloor c \rfloor_\nu
\lfloor \nu \rfloor\! !
}
z^\nu
\label{F43def}.
\eeq
Eq.\ref{F43} is given in ref.\cite{S} in the non-q-deformed case.
Although some q-deformed hypergeometric function properties
are given in this reference, this one is not computed explicitly.
This is why we rederive it rapidly, verifying that the proof of the classical
case extends to the q-deformed case without problem.

The basic idea is to write a finite double sum in two different ways
by inverting the summations:
\beq
\sum_{n=0}^{\infty}
\sum_{p=0}^n
\beta_n \alpha_p u_{n-p}
=
\sum_{p=0}^\infty
\sum_{n=p}^\infty
\beta_n \alpha_p u_{n-p}.
\label{sum}
\eeq
We formally let summations go to infinity for simplicity,
but they are all finite as $\beta_n$ is zero for $n>N$
(it can be checked on the explicit values given below in Eq.\ref{values}).
We define
\beq
S_n=
\sum_{p=0}^n
\alpha_p u_{n-p}
\ ,\
T_p=
\sum_{n=p}^\infty
\beta_n u_{n-p},
\eeq
so that Eq.\ref{sum} reads
\beq
\sum_{n=0}^{\infty}
\beta_n S_n=
\sum_{p=0}^\infty
\alpha_p T_p.
\eeq
Let us  choose
\beq
\alpha_p
\equiv
{
\lfloor x \rfloor_p
\lfloor y \rfloor_p
\over
\lfloor a \rfloor_p
\lfloor p \rfloor\! !
}
\ ,\
\beta_n
\equiv
{
\lfloor z \rfloor_n
\lfloor -N \rfloor_n
\over
\lfloor b \rfloor_n
\lfloor c \rfloor_n
}
\ ,\
u_k
\equiv
{
\lfloor a-x-y \rfloor_k
\over
\lfloor k \rfloor\! !
},
\label{values}
\eeq
so that  the sums $S_n$ and $T_p$
can be computed.
We use identities  of the type
\beq
\lfloor a \rfloor_n
=
1/\lfloor a+n \rfloor_{-n}
\ ,\
\lfloor a \rfloor_n
/
\lfloor a \rfloor_m
=\lfloor a+m \rfloor_{n-m}
\ ,\
\lfloor a \rfloor_n
=
(-1)^n
\lfloor -a-n+1 \rfloor_n,
\label{transf}
\eeq
to which  a meaning
can be given  for any  signs of the integers $n,m$
 --- thanks to the following usual continuation for the products
\beq
\prod_{x=a}^b f(x)
=1/
\prod_{x=b+1}^{a-1} f(x)
\label{prodneg}.
\eeq
One can transform $S_n$ into the following $_3F_2$  hypergeometric
function
\beq
S_n=
{
\lfloor a-x-y \rfloor_n
\over
\lfloor n \rfloor\! !
}
\,_3F_2\left(
^{x,y,-n;}
_{a,x+y-a-n+1;}
1
\right).
\eeq
It is a terminating ($-n$ is a negative integer)
Saalschutzian ($x+y+(-n)+1=a+(x+y-a-n+1)$)
$_3F_2$ series,
and can therefore be summed,  thanks to the q-deformed
Saalschutz theorem given in ref.\cite{S}:
\beq
\,_3F_2\left(
^{x,y,-n;}
_{a,x+y-a-n+1;}
1
\right)
=
{
\lfloor a-x \rfloor_n
\lfloor a-y \rfloor_n
\over
\lfloor a \rfloor_n
\lfloor a-x-y \rfloor_n
}.
\eeq

The same happens to $T_p$ so that Eq.\ref{sum}
reduces to one sum on each side.
These sums can both be put under the form of a $_4F_3$
hypergeometric series,
which finally  gives Eq.\ref{F43}. Q.E.D.

\vskip 3mm

The generalized 6-j coefficient of Eq.\ref{sixj4} can easily
be written with a $_4F_3$ series:
$$
\left\{ ^{j_1}_{j_3}\,^{j_2}_{j_{123}}
\right. \left |^{j_{12}}_{j_{23}}\right\}
=
{
\Xi _{j_2j_3}^{j_{23}}\
\Xi_{{j_1}j_{23}}^{j_{123}}
\over
\Xi_{j_1j_2}^{j_{12}}\
\Xi_{j_{12}j_3}^{j_{123}}
}
{
(-1)^{p_{1,2}}
\lfloor 2j_{23}+1 \rfloor
\lfloor p_{1,23} \rfloor \! !
\lfloor j_{123}+j_{23}-j_{2}-j_{12}+1 \rfloor_{p_{12,3}}
\over
\lfloor p_{12,3} \rfloor \! !
\lfloor p_{1,23}-p_{12,3} \rfloor \! !
\lfloor j_{23}+j_{123}-j_1+1 \rfloor_{p_{1,23}+p_{2,3}+1}
}
$$
\beq
{
\lfloor j_{123}+j_{1}-j_{2}-j_{3}+1 \rfloor_{p_{2,3}}
\over
\lfloor j_1-j_2-j_{12} \rfloor_{p_{12,3}-p_{1,23}}
}
\,_4F_3\left(
^{j_1+j_{23}+j_{123}+2,j_1-j_2-j_{12},-p_{2,3},-p_{12,3};}
_{p_{1,23}-p_{12,3}+1,j_{123}+j_{23}-j_{2}-j_{12}+1,
j_{123}+j_{1}-j_{2}-j_{3}+1;}
1\right),
\label{sixjF1}
\eeq
where we wrote the $_4F_3$ summation index $\nu=y-p_{1,23}$
($\nu$ of Eq.\ref{F43def} and $y$ of Eq.\ref{sixj4}).
The factorial $\lfloor p_{1,23}-p_{12,3} \rfloor\! !$ causes
no problem when
$p_{1,23}-p_{12,3}<0$, since  its vanishing    just compensates
the poles of the hypergeometric function
--- which arise for $\nu\ge p_{12,3}-p_{1,23}$ ---
and cancels the other   terms
($\nu< p_{12,3}-p_{1,23}$)  of the summation,
thereby giving the correct formula.

It is not possible to continue this $_4F_3$ function when transforming
all the spins by $j_i\to - j_i-1$. The reason is that
all the $p_{k,l}$
would  also be transformed into $-p_{k,l}-1$ and
the transformed $_4F_3$ would not have any
negative-integer upper argument. It  would
not terminate and,  possibly,  diverge.
This is why we first transform it using Eq.\ref{F43}.

This hypergeometric series verifies the conditions of Eq.\ref{F43}:
it is terminating
and Saalschutzian.
We can therefore transform it thanks to Eq.\ref{F43}
with
$-N=-p_{2,3},z=j_1-j_2-j_{12},a=p_{1,23}-p_{12,3}+1$
and write it in the strictly equivalent form
$$
\left\{ ^{j_1}_{j_3}\,^{j_2}_{j_{123}}
\right. \left |^{j_{12}}_{j_{23}}\right\}
=
{
\Xi _{j_2j_3}^{j_{23}}\
\Xi_{{j_1}j_{23}}^{j_{123}}
\over
\Xi_{j_1j_2}^{j_{12}}\
\Xi_{j_{12}j_3}^{j_{123}}
}
{
(-1)^{p_{1,2}}
\lfloor 2j_{23}+1 \rfloor
\lfloor p_{1,23} \rfloor \! !
\lfloor j_{12}+j_{123}-j_3+1 \rfloor_{p_{2,3}}
\over
\lfloor p_{12,3} \rfloor \! !
\lfloor p_{1,23}-p_{12,3} \rfloor \! !
\lfloor j_{123}+j_2+j_3-j_1+1 \rfloor_{p_{1,23}+1}
}
$$
\beq
{
\lfloor j_3+j_{123}-j_{12}+1 \rfloor_{p_{12,3}-p_{2,3}}
\over
\lfloor j_1-j_2-j_{12} \rfloor_{p_{12,3}-p_{1,23}}
}
\,_4F_3\left(
^{-j_{12}-j_3-j_{123}-1,j_1-j_2-j_{12},-p_{2,3},p_{1,23}+1;}
_{p_{1,23}-p_{12,3}+1,j_1-j_2-j_3-j_{123},j_{23}-j_{12}-j_2-j_{123};}
1\right).
\label{sixjF2}
\eeq

Although equivalent to the definition Eq.\ref{sixjF1}
for all $p_{k,l}$ positive integers,
this expression Eq.\ref{sixjF2} leads to a different continuation
when $j_i$'s are mapped into $-j_i-1$:
the transformed $_4F_3$ is a well defined series as it
is terminating thanks to its upper
parameter $-p_{1,23}$.
So, the 6-j symbol with spins transformed by $j_i\to - j_i-1$
has a consistent definition.
However, the transformed expression
does not have the same arguments as the defining
formula \ref{sixjF1}.
So, in order to compare them,
we look for an other expression of the 6-j coefficients by transforming
again the definition Eq.\ref{sixjF1} by Eq.\ref{F43}.
This time, we take
$-N=-p_{12,3}$, $z=j_1+j_{23}-j_{123}+1$ and $a=p_{1,23}-p_{12,3}+1$.
Then we make the simple change of variables
$j_1 \iff j_3$, $j_{12} \iff j_{23}$
and get (using the symmetry \ref{symsixj})
$$
\left\{ ^{j_1}_{j_3}\,^{j_2}_{j_{123}}
\right. \left |^{j_{12}}_{j_{23}}\right\}
=
{
\Xi_{j_1j_2}^{j_{12}}\
\Xi_{j_{12}j_3}^{j_{123}}
\over
\Xi _{j_2j_3}^{j_{23}}\
\Xi_{{j_1}j_{23}}^{j_{123}}
}
{
(-1)^{p_{2,3}}
\lfloor 2j_{12}+1 \rfloor
\lfloor p_{12,3} \rfloor \! !
\lfloor j_2+j_3+j_{123}-j_1+2 \rfloor _{p_{1,23}}
\over
\lfloor p_{1,23} \rfloor\! !
\lfloor p_{12,3}-p_{1,23} \rfloor \! !
\lfloor j_{12}+j_{123}-j_3+1 \rfloor_{p_{2,3}+1}
}
$$
\beq
\lfloor j_3+j_{23}-j_2+1 \rfloor _{p_{1,2}-p_{1,23}}
\,_4F_3\left(
^{-p_{1,23},j_3+j_{12}+j_{123}+2,p_{2,3}+1,j_2+j_{12}-j_1+1;}
_{p_{12,3}-p_{1,23}+1,j_2+j_{12}+j_{123}-j_{23}+2,
j_2+j_3+j_{123}-j_1+2;}
1\right).
\label{sixjF3}
\eeq
One already sees that Eq.\ref{sixjF2} transformed
by $j_i\to-j_i-1$ gives
the same $_4F_3$ function as Eq.\ref{sixjF3}.
A little work is required to see that the prefactors
are indeed the same,
in particular,
thanks to Eqs.\ref{transf}, \ref{prodneg},
the factorials transform like
\beq
\lfloor p_{1,23} \rfloor \! !
/
\lfloor p_{12,3} \rfloor \! !
\to
\lfloor -p_{1,23}-1 \rfloor \! !
/
\lfloor -p_{12,3}-1 \rfloor \! !
=
(-1)^{p_{12,3}-p_{1,23}}
\lfloor p_{12,3} \rfloor \! !
/
\lfloor p_{1,23} \rfloor \! !
\eeq
and consequently
$$
{
\Xi_{j_1j_2}^{j_{12}}\
\Xi_{j_{12}j_3}^{j_{123}}
\over
\Xi _{j_2j_3}^{j_{23}}\
\Xi_{{j_1}j_{23}}^{j_{123}}
}
\to
{
\Xi_{-j_1-1\>-j_2-1}^{-j_{12}-1}\
\Xi_{-j_{12}-1\>-j_3-1}^{-j_{123}-1}
\over
\Xi _{-j_2-1\>-j_3-1}^{-j_{23}-1}\
\Xi_{-j_1-1\>-j_{23}-1}^{-j_{123}-1}
}
=
{
\Xi _{j_2j_3}^{j_{23}}\
\Xi_{{j_1}j_{23}}^{j_{123}}
\over
\Xi_{j_1j_2}^{j_{12}}\
\Xi_{j_{12}j_3}^{j_{123}}
}
{
\lfloor 2j_{23}+1 \rfloor
\over
\lfloor 2j_{12}+1 \rfloor
}.
$$
After simplifications, this  proves that
\beq
\left\{ ^{-j_1-1}_{-j_3-1}\,^{-j_2-1}_{-j_{123}-1}
\right. \left |^{-j_{12}-1}_{-j_{23}-1}\right\}
=
\left\{ ^{j_1}_{j_3}\,^{j_2}_{j_{123}}
\right. \left |^{j_{12}}_{j_{23}}\right\}.
\label{sixj-j-1}
\eeq

Finally,
we write this transformation for effective 6-j coefficients,
as they are the ones of interest in fusion or braiding.
We apply the transformation \ref{sixj-j-1} to their
definition \ref{sixjp} and get
\beq
\sixje \Je_1,\Je_3,\Je_2,\Je_{123},\Je_{12},\Je_{23}
=
\sixjxi -\Je_1-1,-\Je_3-1,-\Je_2-1,
-\Je_{123}-1-(\phat_{2,3}+\phat_{1,23})\pi/h,
-\Je_{12}-1-\phat_{1,2}\pi/h,-\Je_{23}-1-\phat_{2,3}\pi/h
\ .
\eeq
This is the definition of an effective 6-j with
entries $-\Je_i-1$.
But this transformation does not treat $\Je_i$ and $\Jehat_i$
symmetrically and will not be convenient for later use.
So, we shift the previous 6-j entries by
\beq
-{\pi\over h}
\left( ^{0}_{1}\,^{0}_{1}
\,^{0}_{1}\right)
-{\pi\over h}
\left( ^{1}_{0}\,^{0}_{1}
\,^{1}_{0}\right)
-{\pi\over h}
\left( ^{0}_{0}\,^{1}_{1}
\,^{1}_{1}\right)
\eeq
which does not change its value thanks to
Eqs.\ref{shift}.
This gives  the more symmetric transformation
\beq
\sixje \Je_1,\Je_3,\Je_2,\Je_{123},\Je_{12},\Je_{23}
=
\sixje -\Je_1-(1+\pi/h),-\Je_3-(1+\pi/h),-\Je_2-(1+\pi/h),
-\Je_{123}-(1+\pi/h),-\Je_{12}-(1+\pi/h),-\Je_{23}-(1+\pi/h),
\label{sixje-j-1}
\eeq
which we use later on.

\section{$\!\!\!\!$STRONG COUPLING REGIME OPERATORS}
\markboth{ 6. Strong coupling operators}
{ 6. Strong coupling operators}
\label{s5}
\subsection{Relation between 6-j symbols}

In this section we restrict ourselves to the central charges
of interest for the strong coupling regime:
$C=1+6(2+s)$.
As explained   in the preliminary  section (2),
the Hilbert spaces to be considered are
${\cal H}^\pm_{s \,  \hbox {\scriptsize phys}}$ defined
by Eqs.\ref{2.16}, \ref{2.17}. The effective
spins, corresponding to Eq.\ref{2.17},   are
\beq
\Bigl\{
\Jme=J+J\pi/h,2J=n+r/(2+s),n\in Z,r=0...1+s
\Bigr\},
\label{h-}
\eeq
with real negative weights,
for ${\cal H}^-_{s \,  \hbox {\scriptsize phys}}$, and
\beq
\Delta(\Jme)=
\Delta(J,J)=
-(2+s)J(J+1),
\label{w-}
\eeq
or
\beq
\Bigl\{
\Jpe=-J-1+J\pi/h,2J=n+r/(2-s),n\in Z,r=0...1-s
\Bigr\},
\label{h+}
\eeq
with real positive weights,
for ${\cal H}^+_{s \,  \hbox {\scriptsize phys}}$, and
\beq
\Delta(\Jpe)=
\Delta(-J-1,J)=
1+(2-s)J(J+1).
\label{w+}
\eeq
As before we define hatted quantities by
$\Jhat^e\,\! ^\pm =J^e\,\!^\pm h/\pi$,
so we have
\beq
\Jmehat=J+Jh/\pi, \quad
\Jpehat=J+(-J-1)h/\pi.
\label{hhat}
\eeq
The aim of this section is a
truncation theorem of the type displayed in  ref.\cite{G3}. Indeed,
we will see  that,
for  these particular values of $C$,
these two subsets of operators are individually closed for
braiding and fusion.

The main tool to prove the closure of the algebra
of the physical operators is that the hatted 6-j coefficients
are equal to unhatted 6-j coefficients up to a sign,
provided $C=1+6(s+2)$, $s=-1,0,1$ if the spins considered
are of the form given in
Eq.\ref{h+} or Eq.\ref{h-}. This type of reasoning was already
presented in ref.\cite{G3}, albeit for different  quantum-group symbols
--- 3-j's, and universal R matrix --- and for half-integer spins only.
Before coming to this result,
we have to prove some preliminary properties.

Using the fact that
\beq
h+\hhat=(C-13)\pi/6=s \pi
\eeq
it is straightforward to prove that for $A$ of the type
\beq
A=n+r/(2+s)
\ ,\
n\in Z,r=0...1+s
\label{frac-}
\eeq
one has ($k\in Z$)
\beq
\lfloorhat A(1+\pi/\hhat)+k \rfloorhat
=
(-1)^{(2+s)(A+k+1)}
\lfloor A(1+\pi/h)+k \rfloor
\label{tr1-}
\eeq
and consequently that ($p\in Z_+$)
\beq
\lfloorhat A(1+\pi/\hhat)+k \rfloorhat_p
=
(-1)^{(2+s)p(A+k+(p+1)/2)}
\lfloor A(1+\pi/h)+k \rfloor_p
\label{tr2-},
\eeq
\beq
\lfloorhat p \rfloorhat \! !
=
(-1)^{sp(p-1)/2}
\lfloor p \rfloor \! !.
\label{tr3}
\eeq
Eqs.\ref{tr2-}, \ref{tr3} allow to prove the following identity
about hypergeometric functions:
\beq
_{k+1}\!\Fhat_k\!
\left(
^{\left(A_i(1+\pi/\hhat)\right)_{i=1...k+1};}
_{\left(A'_i(1+\pi/\hhat)\right)_{i=1...k};}
z
\right)
\!\!
=
_{k+1}\!\! F_k\!
\left(
^{\left(A_i(1+\pi/h)\right)_{i=1...k+1};}
_{\left(A'_i(1+\pi/h)\right)_{i=1...k};}
z(-1)^{\left(\sum(sn_i+r_i)-\sum(sn'_i+r'_i)+1\right)}
\right)
\label{tr4-}
\eeq
with the $k+1$ upper entries and $k$ lower entries obtained
from $A_i$ and $A'_i$ of the type \ref{frac-}:
$$
A_i=n_i+r_i/(2+s)
,n_i\in Z,r_i=0...1+s
\ ;\
A'_i=n'_i+r'_i/(2+s)
,n'_i\in Z,r'_i=0...1+s.
$$
There is a similar property for hypergeometric function
with arguments including no ``$\pi/h$ part'',
but they are then restricted to be integers\cite{G3}.
So, it is clear that if we want to be able to relate
hatted and unhatted 6-j coefficients for
fractionnal spins
we have to consider effective spins
introduced in \ref{h-} (and \ref{h+}) only.
We shall generically note $\Jme_i=J_i+J_i\pi/h$
and $\Jpe_i=-J_i-1+J_i\pi/h$
the spins of the type
\ref{h-} and \ref{h+} respectively.

Consider first the hatted part of
the effective 6-j coefficients,
using their definition \ref{sixjp}:
\beq
\sixjehat \Jmehat_1,\Jmehat_3,\Jmehat_2,
\Jmehat_{123},\Jmehat_{12},\Jmehat_{23}
=
\sixjxihat J_1+J_1\pi/\hhat,J_3+J_3\pi/\hhat,
\qquad J_2+J_2\pi/\hhat,J_{123}+(J_1+J_2+J_3)/\pi/\hhat,
J_{12}+(J_1+J_2)\pi/\hhat,{J_{23}+(J_2+J_3)\pi/\hhat}
\ .
\eeq
We insert  the expression \ref{sixjF1} of the 6-j in terms
of $_4F_3$ function and see that its arguments are
precisely of the type $A_i(1+\pi/h)$.
So, using  Eq.\ref{tr4-} for the $_4F_3$ function
and Eqs.\ref{tr2-}, \ref{tr3} for the
prefactors,
we notice that Eq.\ref{tr4-} yields
no extra sign for the argument $z$
in this case,  as 6-j symbols involve Saalschutzian $_4F_3$ functions.
Thus we  get
$$
\sixjehat \Jmehat_1,\Jmehat_3,\Jmehat_2,\Jmehat_{123},\Jmehat_{12},\Jmehat_{23}
=
$$
\beq
(-1)^{(2+s)
\phi(J_1,J_2,J_3,J_{12},J_{23},J_{123})
}
\sixjxi J_1+J_1\pi/h,J_3+J_3\pi/h,
J_2+J_2\pi/h,J_{123}+(J_1+J_2+J_3)/\pi/h,
J_{12}+(J_1+J_2)\pi/h,J_{23}+(J_2+J_3)\pi/h
\eeq
with
$$
\phi(J_1,J_2,J_3,J_{12},J_{23},J_{123})=
p_{1,2}(2J_2+2J_3)+p_{2,3}2J_3+p_{1,23}2J_{23}+p_{12,3}2J_3
$$
\beq
+p_{1,2}(p_{1,2}+1)/2
+p_{2,3}(p_{2,3}+1)/2
+p_{1,23}(p_{1,23}+1)/2
+p_{12,3}(p_{12,3}+1)/2
\label{phi}
\eeq
with $p_{k,l}\equiv j_k+j_l-j_{kl}$ as usual.
In terms of effective 6-j coefficients, it reads
\beq
\sixjehat \Jmehat_1,\Jmehat_3,\Jmehat_2,\Jmehat_{123},\Jmehat_{12},\Jmehat_{23}
=
(-1)^{(2+s)
\phi(J_1,J_2,J_3,J_{12},J_{23},J_{123})
}
\sixje \Jme_1,\Jme_3,\Jme_2,\Jme_{123},\Jme_{12},\Jme_{23}
\label{sixjeh-}.
\eeq

In the argument just given, the hatted spins are obtained from
the unhatted ones
 by exchanging $h$ and $\hhat$. This  is not true for the other
case to which we come next. According to Eq.\ref{hhat}, exchanging
$h$ and $\hhat$ in $\Jpe$ does not give $\Jpehat$, but rather
$-(J+1)+J h/\pi\equiv -\Jpe-1-\pi/h$
which is not the same  in general.
We shall see that the transformations of the type
 \ref{tr4-} will still correspond to
exchanging $h$ and $\hhat$
and not $\Je$ and $\Jehat$
(one could not distinguish between them in the $\Jme$ case).
It is one  novelty of  the positive-weight spin case
to which we are coming now.
Another    is that we will have to use the results  of last section,
as the $\Jpe_i$'s  involve negative spins and more fundamentally,
since  the unhatted 6-j coefficients  with spins $\Jpe_i$ involve negative
screening-numbers $p_{k,l}$.
 The final result will be a relation,  where $\Jpe$
is transformed into $\Jpehat$,  similar to the previous case.
The counterpart of Eqs.\ref{frac-}--\ref{tr4-} is now that, for
\beq
B=n+r/(2-s)
\ ,\
n\in Z,r=0...1-s
\label{frac+}
\eeq
the useful identities are
\beq
\lfloorhat B(1-\pi/\hhat)+k \rfloorhat
=
(-1)^{(2-s)(B+k+1)}
\lfloor B(1-\pi/h)+k \rfloor,
\label{tr1+}
\eeq
\beq
\lfloorhat B(1-\pi/\hhat)+k \rfloorhat_p
=
(-1)^{(2-s)p(B+k+(p+1)/2)}
\lfloor B(1-\pi/h)+k \rfloor_p
\label{tr2+}
\eeq
and
\beq
_{k+1}\!\Fhat_k\!
\left(
^{\left(B_i(1-\pi/\hhat)\right)_{i=1...k+1};}
_{\left(B'_i(1-\pi/\hhat)\right)_{i=1...k};}
z
\right)
\!\!
=
_{k+1}\!\!\! F_k\!
\left(
^{\left(B_i(1-\pi/h)\right)_{i=1...k+1};}
_{\left(B'_i(1-\pi/h)\right)_{i=1...k};}
z(-1)^{\left(\sum(sn_i+r_i)-\sum(sn'_i+r'_i)+1\right)}
\right)
\label{tr4+}
\eeq
with
$$
B_i=n_i+r_i/(2-s)
,n_i\in Z,r_i=0...1-s
\ ;\
B'_i=n'_i+r'_i/(2-s)
,n'_i\in Z,r'_i=0...1-s.
$$
As said above, it turns out that the transformation \ref{tr4+}
allows to relate unhatted 6-j coefficients with hatted 6-j
coefficients with symmetrized spins.
By definition
\beq
\sixje \Jpe_1,\Jpe_3,\Jpe_2,\Jpe_{123},\Jpe_{12},\Jpe_{23}
\equiv
\sixjxi -J_1-1+J_1\pi/h,-J_3-1+J_3\pi/h,
\qquad -J_2-1+J_2\pi/h,-J_{123}-1+(J_1+J_2+J_3)/\pi/h,
-J_{12}-1+(J_1+J_2)\pi/h,{-J_{23}-1+(J_2+J_3)\pi/h}
\eeq
and, like in the negative weight case,
using Eqs.\ref{tr3}, \ref{tr2+}, \ref{tr4+},
it can be transformed into
$$
(-1)^{(2-s)
\phi(J_1,J_2,J_3,J_{12},J_{23},J_{123})
}
\sixjxihat -J_1-1+J_1\pi/\hhat,-J_3-1+J_3\pi/\hhat,
\qquad -J_2-1+J_2\pi/\hhat,-J_{123}-1+(J_1+J_2+J_3)\pi/\hhat,
-J_{12}-1+(J_1+J_2)\pi/\hhat,{-J_{23}-1+(J_2+J_3)\pi/\hhat}
$$
or, in terms of effective spins and effective 6-j's
\beq
\sixje \Jpe_1,\Jpe_3,\Jpe_2,\Jpe_{123},\Jpe_{12},\Jpe_{23}
\!\!
=
\!
(-1)^{(2-s)
\phi(J_1,J_2,J_3,J_{12},J_{23},J_{123})}
\sixjehat -\Jpehat_1-1-\pi/\hhat,-\Jpehat_3-1-\pi/\hhat,-\Jpehat_2-1-\pi/\hhat,
-\Jpehat_{123}-1-\pi/\hhat,-\Jpehat_{12}-1-\pi/\hhat,-\Jpehat_{23}-1-\pi/\hhat
\label{sixjehat}
\eeq
following the remark above that the symmetrized spins
$-J_i-1+J_i\pi/\hhat$ can be written $-\Jpehat_i -1-\pi/\hhat$.
Then, the result of last section, Eq.\ref{sixje-j-1},
makes the link with positive screening number 6-j symbols and gives
\beq
\sixje \Jpe_1,\Jpe_3,\Jpe_2,\Jpe_{123},\Jpe_{12},\Jpe_{23}
=
(-1)^{(2-s)
\phi(J_1,J_2,J_3,J_{12},J_{23},J_{123})
}
\sixjehat \Jpehat_1,\Jpehat_3,\Jpehat_2,\Jpehat_{123},\Jpehat_{12},\Jpehat_{23}
\label{sixjeh+}.
\eeq
In this case of the positive weight spins,
we transformed the unhatted effective spins and 6-j's
into their hatted counterparts,
whereas in the negative weight case we made the the contrary.
It is because the hatted 6-j symbols are the ones with
positive screening numbers.

We would like to emphasize that the compact notation
should not hide the very different natures of the transformations
\ref{sixjehat} and \ref{sixje-j-1}
which are necessary to obtain Eq.\ref{sixjeh+}.
The transformation \ref{sixje-j-1} relates 6-j coefficients with
positive number of
screening charges
with 6-j coefficients with
negative
number of
screening charges,
defining the latter by a suitably chosen continuation
of $_4F_3$ function.
It is valid for generic central charge and spins.
The transformation \ref{sixjehat} relates hatted and unhatted
6-j coefficients with the same type (positive or negative)
of screening charge number.
It involves no continuation but is only valid
for $C=1+6(2+s)$ and spins
$\Jpe_i$ of the type \ref{h+}.

\vskip 3mm
\subsection{The physical fields  with negative conformal weights}

The necessary tools are ready now and we build the physical operators.
The sign $(2\pm s)\phi(J_i)$ that shows up in
Eqs.\ref{sixjeh-}, \ref{sixjeh+} can be grouped in four identical terms
for the four vertices of fusion and an extra fifth term.
It is therefore natural to include the four terms in the definition
of the operators in order to get rid of them and be able to use
the orthogonality of the 6-j coefficients to prove the closure
of the algebra.
So, we define
\beq
{\cal P}_{\Jme_{12}}\chi_-^{(J_1)}
\equiv
\sum_{J_2,p_{1,2}\in Z_+}
(-1)^{(2+s)(2J_2p_{1,2}
+{p_{1,2}(p_{1,2}+1)\over 2})}
g_{\Jme_1\Jme_2}^{\Jme_{12}}
{\cal P}_{\Jme_{12}}
V^{(\Jme_1)}
{\cal P}_{\Jme_2}
\label{chi-}
\eeq
with $p_{1,2}\equiv J_1+J_2-J_{12}$.
It can as well be summed over $\Jme_{12}$
and written,
using the closure relation in the Hilbert space for the l.h.s.,
\beq
\chi_-^{(J_1)}
\equiv
\sum_{J_{12},J_2,p_{1,2}\in Z_+}
(-1)^{(2+s)(2J_2p_{1,2}
+{p_{1,2}(p_{1,2}+1)\over 2})}
g_{\Jme_1\Jme_2}^{\Jme_{12}}
{\cal P}_{\Jme_{12}}
V^{(\Jme_1)}
{\cal P}_{\Jme_2}
\label{chidef-}.
\eeq

The fusion of such operators is obtained from Eq.\ref{fusion}:
$$
\chi_-^{(J_1)}
\chi_-^{(J_2)}
=
\sum_{J_{123},J_{23},J_3}
\!\!\!
(-1)^{(2+s)(2J_{23}p_{1,23}+2J_3p_{2,3}+{p_{2,3}(p_{2,3}+1)\over 2}
+{p_{1,23}(p_{1,23}+1)\over 2})}
$$
\beq
g_{\Jme_2\Jme_3}^{\Jme_{23}}
g_{{\Jme_1}\Jme_{23}}^{\Jme_{123}}
{\cal P}_{\Jme_{123}}
V^{(\Jme_1)}
{\cal P}_{\Jme_{23}}
V^{(\Jme_2)}
{\cal P}_{\Jme_3}
\label{chichi}
\eeq
$$
=
\sum_{J_{123},J_3}
\sum_{J_{23},\Je_{12}}
(-1)^{(2+s)(2J_{23}p_{1,23}+2J_3p_{2,3}+{p_{2,3}(p_{2,3}+1)\over 2}
+{p_{1,23}(p_{1,23}+1)\over 2})}
g_{\Jme_1\Jme_2}^{\Je_{12}}
g_{\Je_{12}\Jme_3}^{\Jme_{123}}
$$
\beq
\sixje \Jme_1,\Jme_3,\Jme_2,\Jme_{123},\Je_{12},\Jme_{23}
\sixjehat \Jmehat_1,\Jmehat_3,\Jmehat_2,\Jmehat_{123},\Jehat_{12},\Jmehat_{23}
\sum _{\{\nu_{12}\}}
{\cal P}_{\Jme_{123}}
V^{(\Je_{12},\{\nu_{12}\})} {\cal P}_{\Jme_{3 }}
<\!{\Je_{12}},{\{\nu_{12}\}} \vert
V ^{(\Jme_1)} \vert {\Jme_2} \! >.
\label{fus1}
\eeq
It must be noticed that we wrote $\Je_{12}(=J_{12}+\Jhat_{12}\pi/h)$
and not $\Jme_{12}(=J_{12}+J_{12}\pi/h)$
as, of course, the fusion a priori  creates operators of the full algebra
($J_{12}$ and $\Jhat_{12}$ are different)
and not of the type displayed on Eq.\ref{h-}.
So, the sum is for $J_1+J_{23}-J_{123},J_2+J_3-J_{23}\in Z_+$
and $\Jme_1+\Jme_2-\Je_{12},\Je_{12}+\Jme_3-\Jme_{123}\in Z_++\pi/hZ_+$.
However, the definition of the effective 6-j coefficients
Eq.\ref{sixjp} allows us to transform them in the following way:
$$
\sixje \Jme_1,\Jme_3,\Jme_2,\Jme_{123},\Je_{12},\Jme_{23}
=
\sixje \Jme_1,\Jme_3,\Jme_2,\Jme_{123},J_{12}+J_{12}\pi/h,\Jme_{23}
$$
and
$$
\sixjehat \Jmehat_1,\Jmehat_3,\Jmehat_2,\Jmehat_{123},\Jehat_{12},\Jmehat_{23}
=
\sixjehat \Jmehat_1,\Jmehat_3,\Jmehat_2,\Jmehat_{123},
\Jhat_{12}+\Jhat_{12}\pi/\hhat,\Jmehat_{23}
$$
$$
=
(-1)^{(s+2)
\phi(J_1,J_2,J_3,\Jhat_{12},J_{23},J_{123})
}
\sixje \Jme_1,\Jme_3,\Jme_2,\Jme_{123},\Jhat_{12}+\Jhat_{12}\pi/h,\Jme_{23}
$$
where the last equation comes from Eq.\ref{sixjeh-}.
Putting this in \ref{fus1}
and simplifying the signs we get
$$
\sum_{J_{123},J_3,J_{12}}
\sum_{\Jhat_{12}}
(-1)^{(2+s)(2\Jhat_3(\Jhat_{12}+J_3-J_{123})+(2J_2+2J_3)(J_1+J_2-\Jhat_{12})
{(J_1+J_2-\Jhat_{12})(J_1+J_2-\Jhat_{12}+1)\over 2})}
$$
$$
(-1)^{(s+2)
{(\Jhat_{12}+J_3-J_{123})(\Jhat_{12}+J_3-J_{123}+1)\over 2}
}
\!\!
\left(
\sum_{J_{23}}
\sixje \Jme_1,\Jme_3,\Jme_2,\Jme_{123},J_{12}+J_{12}\pi/h,\Jme_{23}
\sixje \Jme_1,\Jme_3,\Jme_2,\Jme_{123},\Jhat_{12}+\Jhat_{12}\pi/h,\Jme_{23}
\right)
$$
\beq
g_{\Jme_1\Jme_2}^{\Je_{12}}
g_{\Je_{12}\Jme_3}^{\Jme_{123}}
\sum _{\{\nu_{12}\}}
{\cal P}_{\Jme_{123}}
V^{(\Je_{12},\{\nu_{12}\})} {\cal P}_{\Jme_{3 }}
<\!{\Je_{12}},{\{\nu_{12}\}} \vert
V ^{(\Jme_1)} \vert {\Jme_2} \! >.
\label{fus2}
\eeq
The sum over $J_{23}$ can be performed
as all the dependence in $J_{23}$ is in the 6-j coefficients.
It is precisely the orthogonality relation
which yields $\delta_{J_{12},\Jhat_{12}}$
and supresses the sum over $\Jhat_{12}$.
It is easy to see that thanks to the signs and the $g$ coefficients
it gives back $\chi$ operators.
We write the result after removal of the sums over
$\Jme_{123}$ and $\Jme_3$:
$$
\chi_-^{(J_1)}
\chi_-^{(J_2)}
{\cal P}_{\Jme_3}
=
\sum_{J_{12}<J_1+J_2}
(-1)^{(2+s)2J_3(J_1+J_2-J_{12})}
$$
\beq
\sum_{\{\nu_{12}\}}
\chi_-^{(J_{12},\{\nu_{12}\})}
{\cal P}_{\Jme_3}
<\!{\Jme_{12}},{\{\nu_{12}\}} \vert
\chi_-^{(J_1)} \vert {\Jme_2} \! >
\label{fuschi-}.
\eeq
We kept the projector ${\cal P}_{\Jme_3}$ at the right of the operators,
as the value of the spin at the right is necessary to write
the sign $(2+s)2J_3(J_1+J_2-J_{12})$.
But we could equivalently sum over $J_3$
and use the operator $\varpi$
of eigenvalues $\varpi_{\Jme}=\varpi_0+2\Jme$
(which does not commute with the vertex operators):
\beq
\chi_-^{(J_1)}
\chi_-^{(J_2)}
=
\!\!\!
\sum_{J_{12}<J_1+J_2,\{\nu_{12}\}}
\!\!\!\!\!\!\!\!\!
\chi_-^{(J_{12},\{\nu_{12}\})}
(-1)^{(1+h/\pi)(\varpi-\varpi_0)(J_1+J_2-J_{12})}
<\!{\Jme_{12}},{\{\nu_{12}\}} \vert
\chi ^{(J_1)} \vert {\Jme_2} \! >
\label{fuschi2-}.
\eeq

The closure of the braiding works in the same way.
The only differences are in the signs,
in particular
the extra phase of the braiding Eq.\ref{braiding}.
The braiding of the two $\chi$ fields of Eq.\ref{chichi}
involves the 6-j coefficients
$$
\sixje \Jme_1,\Jme_2,\Jme_3,\Jme_{123},\Jme_{13},\Jme_{23}
\sixjehat \Jmehat_1,\Jmehat_2,\Jmehat_3,\Jmehat_{123},\Jmehat_{13},\Jmehat_{23}
$$
which yield the sign
$
(-1)^{(s+2)
\phi(J_1,J_3,J_2,J_{13},J_{23},J_{123})}.
$
The signs combine correctly thanks to the following identity
$$
(s+2)\Bigl [
p_{1,3}(2J_3+2J_2)+p_{3,2}2J_2+p_{1,23}2J_{23}+p_{13,2}2J_2
\Bigr ]
$$
$$
+\epsilon \Bigl [
\Delta(\Jme_{13})+
\Delta(\Jme_{23})-
\Delta(\Jme_{3})-
\Delta(\Jme_{123})
\Bigr ]
=
$$
\beq
(s+2)\Bigl [
p_{1,23}2J_{23}+p_{2,3}2J_3+p_{2,13}2J_{13}+p_{1,3}2J_3
-2\epsilon J_1J_2
\Bigr ]
\hbox{ (mod 2)}
\label{sgnbrd}.
\eeq
Using this we prove
\beq
\chi_-^{(J_1)}
\chi_-^{(J_2)}
=
e^{-2i\pi\epsilon(2+s)J_1J_2}
\chi_-^{(J_2)}
\chi_-^{(J_1)}
\label{brdchi-}.
\eeq

\vskip 3mm

It is easy to check that the $\chi$ operators verify the polynomial
equations.
For example,
the orthogonality is satisfied thanks to the $\epsilon$ sign
in the braiding Eq.\ref{brdchi-}.
The pentagonal identity expressing the associativity
of
$$
\chi_-^{(J_1)}
\chi_-^{(J_2)}
\chi_-^{(J_3)}
=
\sum_{J_4,J_{34},J_{234},J_{1234}}
{\cal P}_{\Jme_{1234}}
\chi_-^{(J_1)}
{\cal P}_{\Jme_{234}}
\chi_-^{(J_2)}
{\cal P}_{\Jme_{34}}
\chi_-^{(J_3)}
{\cal P}_{\Jme_{4}}
$$
works thanks to
$$
(2+s)
\Bigl [
(J_2+J_3-J_{23})2J_4
+(J_1+J_{23}-J_{123})2J_4
+(J_1+J_2-J_{12})2J_3
\Bigr ]
$$
$$
=
(s+2)
\Bigl [
(J_1+J_2-J_{12})2J_{34}
+
(J_{12}+J_3-J_{123})2J_4
\Bigr ]
\hbox{ (mod 2)}
{}.
$$

\subsection{The physical fields with positive conformal weights}

The operators with positive weights are build similarly.
They involve
negative screening numbers and thus
required some more work in last section,
but now that the formula \ref{sixjeh+}
similar to \ref{sixjeh-} relating hatted
and unhatted 6-j's has been worked out,
they can be built along the same line.
We define
\beq
\chi_+^{(J_1)}
\equiv
\sum_{J_{12},J_2,p_{1,2}\equiv J_1+J_2-J_{12}\in Z_+}
(-1)^{(2-s)(2J_2p_{1,2}
+{p_{1,2}(p_{1,2}+1)\over 2})}
g_{\Jpe_1\Jpe_2}^{\Jpe_{12}}
{\cal P}_{\Jpe_{12}}
V^{(\Jpe_1)}
{\cal P}_{\Jpe_2}
\label{chidef+}.
\eeq
The closure of fusion works similarly and gives
$$
\chi_+^{(J_1)}
\chi_+^{(J_2)}
{\cal P}_{\Jme_3}
=
\sum_{J_{12}<J_1+J_2}
(-1)^{(2-s)2J_3(J_1+J_2-J_{12})}
$$
\beq
\sum_{\{\nu_{12}\}}
\chi_+^{(J_{12},\{\nu_{12}\})}
{\cal P}_{\Jme_3}
<\!{\Jpe_{12}},{\{\nu_{12}\}} \vert
\chi ^{(J_1)} \vert {\Jpe_2} \! >
\label{fuschi+}.
\eeq
There is a slight change concerning the braiding,
as the weights are not the same.
The positive weights of Eq.\ref{w+}
can be obtained from the negative ones of Eq.\ref{w-}
by a change of sign and a change
of $s$ into $-s$ (up to an extra 1).
Thus we get the right formula from Eq.\ref{sgnbrd}
by changing $s$ and $\epsilon$ in their opposite,
and finally we have for braiding
\beq
\chi_+^{(J_1)}
\chi_+^{(J_2)}
=
e^{2i\pi\epsilon(2-s)J_1J_2}
\chi_+^{(J_2)}
\chi_+^{(J_1)}
\label{brdchi+}.
\eeq
Finally, we note that we could have chosen
\beq
\Bigl\{
\Jpe=J+(-J-1)\pi/h,2J=n+r/(2-s),n\in Z,r=0...1-s
\Bigr\},
\label{hb+}
\eeq
instead of Eq.\ref{h+}. This amounts to  exchanging
$h$ and $\hhat$ everywhere in the above formulae,
and the discussion is the same.
This possibility will be important in the coming section.
This other possibility may be deduced  very simply from the above,
by noting that  one only needs to change $J$ into $-J-1$ everywhere.
This leads to the replacement of Eq.\ref{chidef+} by
\beq
\chi_+^{(J_1)}
\equiv
\sum_{J_{12},J_2,p_{1,2}\in Z_+}
(-1)^{(2-s)(2J_2(p_{1,2}+1)
+{p_{1,2}(p_{1,2}+1)\over 2})}
g_{\Jpe_1\Jpe_2}^{\Jpe_{12}}
{\cal P}_{\Jpe_{12}}
V^{(\Jpe_1)}
{\cal P}_{\Jpe_2}
\label{chidef+p}.
\eeq

\subsection{More about the half-integer case  }

As we saw in section \ref{s2} for the 6-j coefficients
and in section \ref{s3} for the operators,
it is possible to extend these algebras to continuous
spins provided the full triangular inequalities (TI3)
are replaced by  TI1 (one inequality per vertex).
These TI1's  are the ``selection rules'' we used
for our vertex operators to build the physical $\chi$ operators.
But then, a question arises:
what happens when we use TI1  and that some spins
happen ``accidentally'' to be half-integers?

When deriving the general algebra in sections \ref{s2} and \ref{s3}
we could think that this case was of probability zero
and we did not ask this question.
But in the case of our operators of the strong coupling
regime with spins $J_i=(n+r/(2\pm s))/2$,
this is very likely to happen.
For example by  fusing the  spins $J_1=5/6$ and $J_2=1/6$
we get $J_{12}=1,0,-1...$.
The questions  are   then:
Are there singularities arising?
Are we entitled to go to negative $J_{12}$ in such a case
as is allowed by TI1?
Or, on the contrary, should we consider TI3 (or simply TI2)
in such a case?....

The answer can be found in the fact that the polynomial
equations,  being   rational,
always guarantee  the consistency of the algebra.
There are three different algebras with TI3, TI2 or TI1.
Once one of these three possibilities is chosen
the algebra is consistent, but, if one of the spins
that was not forseen to be half-integer when making this choice
happens ``accidentally'' to be half-integer,
there are zeros and singularities arising.
They cancel however in the polynomial equations
due to their rational character.
In particular,
this is the case in the traditional construction of Gervais et al
where the zero-mode of the incoming and outgoing states
is continuous, so that one is in the case TI2.
It can easily be checked,
 --- even for basic $J=1/2$ operators, the braiding and fusion  of which
are directly deduced from the differential equation ---
that there are zeros and singularities arising
 if  the zero-mode $\varpi$  ever happens to be equal to
$\varpi_0+2J$, with $2J$ integer.
 At the level of 6-j's,
one can either keep TI2 and have singularities for some discrete
values of the zero-mode,
or use TI3 and have perfectly finite results,
as was done in ref.\cite{CGR1}.

In the strong coupling case, however, we must consider
 spins of the form  $J_i=(n+r/(2\pm s))/2$
in order to complete modular invariance\cite{GR}.
So we have to use TI1.
And half-integer spins arise naturally from non-half-integer ones.
Hence,
we need such vertex operators like
\beq
\epsffile{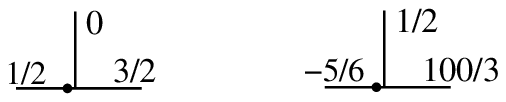}
\label{sing}.
\eeq
Although surprising when one is familiar with TI3,
they can indeed be constructed.
The first one is simply a screening operator
and the second one has got many extra screening operators.
They are not considered usually as they are singular.
What saves us in this case of strong coupling regime
is that the singularities of the operators that we must consider
cancel each other thanks to the particular choice of $J_i$ and $\Jhat_i$.
More precisely,
it means that although the 6-j coefficients
considered all along this section,
and in particular in Eq.\ref{fus2},
may be singular or zero,
nevertheless,
the orthogonality relation Eq.\ref{orthog} holds,
as it is an equation between rational functions,
and it allows to go from Eq.\ref{fus2} to Eq.\ref{fuschi-},
cancelling the singularities\footnote{
There is however a little subtelty in the fact that the equations
\ref{sixjeh-}, \ref{sixjeh+} relating hatted and unhatted 6-j
coefficients are not identities at the level of rational fractions
but for the particular values \ref{h-} and \ref{h+} of the spins.}.
This is one more miracle of this construction
for the strong coupling regime:
although for the basic $V$ operators the
choice of spins \ref{h-} or \ref{h+}
leads to singularities
coming from the 6-j coefficients (until Eq.\ref{fus2}),
these singularities disappear for the physical $\chi$ operators
in Eqs.\ref{fuschi-}, \ref{brdchi-}, \ref{fuschi+}, \ref{brdchi+}.

We want to point out that it has some astonishing implications,
like the following:
the orthogonality Eq.\ref{orthog} can be used
with spins $j_1$, $j_2$, $j_3$, $j_{123}$ half-integers such that
it is possible to find spins $j_{12}$ and $j_{23}$ such that 4 TI3's
are satisfied.
However, we sum over $j_{23}$ verifying TI1 and thanks to the
cancellation of zeros and singularities the result
is still the identity matrix for a range of $j_{12}$
and $j'_{12}$ determined by TI1: $j_{12},j'_{12}\in [j_{123}-j_3,j_1+j_2]$.
It seems curious as for the same spins
there is as well a standard orthogonality relation
with 4 TI3's.
There is no incompatibility between both of them:
if instead, we chose to sum over $j_{23}$ with TI3 i.e.
$|j_{123}-j_1|,|j_2-j_3|\le j_{23}\le j_2+j_3,j_1+j_{123}$,
the result is now (even for $j_{12}$
and $j'_{12}$ verifying TI1, hence in $[j_{123}-j_3,j_1+j_2]$)
the projector on the space of spins  $j_{12}$ and $j'_{12}$
verifying TI3:
$j_{12},j'_{12}\in$
[max($|j_{123}-j_3|,|j_1-j_2|$),min($j_2+j_3,j_{123}+j_3$)].
This is the standard orthogonality relation.
So, if the spins are half-integers,
we can consistently choose either to stay in the subspace of
vertices verifying TI3, or to go to the larger space
of vertices verifying TI1 (and have singularities
in the case of $V$ operators).
Once this choice is made on the initial vertex operators,
the vertex operators obtained by braidings and fusions remain
of the same kind.
The same is true with the intermediate case of TI2.
All this was checked numerically on many exemples.

Since  TI1 must be the selection rules
in every case here,
the fusion Eqs.\ref{fuschi-}, \ref{fuschi2-}, \ref{fuschi+}
involve an infinite sum as $J_{12}$ is not bounded from below.
This is not different from the braiding Eqs.\ref{brdchi-}, \ref{brdchi+},
where the intermediate states
between $\chi^{(J_2)}$ and $\chi^{(J_1)}$ are in infinite number.
However, if all the four external spins are specified (i.e.
if the left and right states are specified as well),
these infinite sums disappear.

The same question must be asked concerning the $g$ coefficients.
The answer is the same:
they are finite in our case in contrast with the general case.
 More precisely,
one can see that in the general case, if  some spins
ever turn out to be half-integer whereas the full
triangular inequalities are not fulfilled,
there are zeros and poles arising.
They do not cancel with the ones of the 6-j coefficients
in general (see e.g. the fusion coefficient for spin 1/2
Eq.\ref{fus1/2}).
However, a careful analysis,
that we do not detail here for brevity sake,
shows that the $g$'s of interest here,
i.e. with all spins such that $\Jhat=J$ or $\Jhat=-J-1$,
never bring poles
(to see this, use the expression of $g$ in terms of arbitrary
path Eq.\ref{H2},
note that there are only zeros or poles when changing of
quarter of plane,
and that in both of our cases,
the beginning and ending points of the paths are on the same diagonal,
which involves two changes of quarter plane that cancel one an other).
They are never zero except when one of the three spins
involved in a $g$ coefficient is $J=-1/2, \Jhat=-J-1=-1/2$.
This agrees with the leg-factors of the final result Eq.\ref{7.26}
which are zero in such a case only.
This is not really  surprising
as the representation of spin -1/2 is very peculiar.

\section{TOPOLOGICAL MODELS}
\markboth{7.  Topological models}{7.  Topological models}

\subsection{The vertex}
As already indicated, we may consider two copies of the strongly coupled
models under consideration with central charges
\beq
C=1+6(s+2), \quad c=1+6(-s+2).
\label{7.1}
\eeq
The second plays the role of matter, while the first may be considered as
gravity. The latter gives a proper dressing,  since obviously,
\beq
C+c=26
\label{7.2}.
\eeq
In this section, we determine the corresponding three-point functions.
Let us first establish the framework. Since it is very close to the
one already put forward for the weak coupling regime (see ref.\cite{G5}),
we shall be rather brief.

First, we shall be discussing closed surfaces, so that we should include
both left and right movers. As explained in ref.\cite{G5}, the discussion
carried out so far only applies  to the holomorphic components, which are
functions of $z=\tau+i \sigma$ ($\tau$
 is the time coordinate on the cylinder).
For the antiholomorphic components, one should
change $i$ into $-i$ everywhere. Thus if we call $\chib_\pm^{(J)}$, the
corresponding physical fields,  the braiding relations Eqs.\ref{brdchi-},
\ref{brdchi+} are changed into
\beq
\chib_\pm^{(J_1)}(\tau-i \sigma_1)\> \chib_\pm^{(J_2)}(\tau-i \sigma_2)=
e^{2\pi i(2\mp s)J_1 J_2  \epsilon(\sigma_1-\sigma_2)}\>
\chib_\pm^{(J_2)}(\tau-i \sigma_2)\>  \chib_\pm^{(J_1)}(\tau-i \sigma_1).
\label{7.3}
\eeq
Since the braiding of the holomorphic components is a simple
phase, we construct\footnote{As usual  we take the
two holomorphic components to commute.}
local fields  simply from  products of the
form
  $\chi_\pm^{(J)}(\tau+i \sigma)
\chib_\pm^{(\Jb)}(\tau-i \sigma)$
with suitable  $J$, and $\Jb$.
We shall use them   to describe
the coupling of strongly coupled  gravity. They are
the generalizations   of the exponentials of the Liouville
field,  to  the present situation.
Second the matter (with central charge $c$) is treated much like gravity.
Since by construction, $c$ belongs to the set of special values,
there exist
physical fields noted $\chip_\pm^{(J)}(\tau+i \sigma)$, and
$\chibp_\pm^{(J)}(\tau-i \sigma)$ similar to the above. The quantum
group structure has parameters $h'$, and $\hhat'$ (we systematically
distinguish symbols related with matter by  a prime), such that
\beq
h+\hhat'=\hhat+h'=0.
\label{7.4}
\eeq
We consider only the simplest solution of the
BRST cohomology, and  take  vertex operators of the form
\beq
{\cal V}_{J', \Jb'}= \chi_+^{(J')}\>  \chib_+^{(\Jb')}\>  \chip_-^{(J')}\>
\chibp_-^{(\Jb')}.
\label{7.5}
\eeq
It follows from Eq.\ref{2.8}
that ${\cal V}_{J' \Jb'}$ has conformal weights $(1,1)$
as required. At this point let us note a basic difference with the
weak coupling regime. There the Liouville field only involves
left and right movers with equal quantum group spins. Thus, only
three-point functions with $J'=\Jb'$ are considered\cite{G5}. Here, the
quantum spins of the two holomorphic components are not related so far,
and will not be set equal
 (more about this below).
The values of the quantum group spins which appear in the last equation
may be seen on Eqs.\ref{chidef-} and \ref{chidef+}. Concerning the
latter, we have to make an important point. As noted at the end
of the previous section, there are two ways to define the fields $\chi_+$,
which correspond to the existence of the two screening charges.  For
physical reasons, which we shall spell out below, we shall take one choice
(Eq.\ref{h+}) for the holomorphic components, and the
other (Eq.\ref{hb+}) for the antiholomorphic one.
Thus we let
\beqa
\Je\equiv \Jpe&=&(-J'-1)+J'  \pi/h,
\quad \Jep\equiv \Jep^-=J'+J' \pi/h' = J'(1- h/\pi),\nnn
\Jeb\equiv \Jeb^+&=&\Jb'+(-\Jb'-1) \pi/h,
\quad \Jebp\equiv \Jebp^-=\Jb'+\Jb' \pi/h'= \Jb'(1-h/\pi).
\label{7.6}
\eeqa
As just noted,  the
upper indices $\pm$ are omitted   from now on.
  Concerning the Hilbert spaces, our previous discussion
shows that the fields $\chi_+$ (resp. $\chi_-$)   have a consistent
restriction to
the physical  Hilbert space
${\cal H}_{s \,  \hbox {\scriptsize phys}}^+$
(resp. ${\cal H}_{s \,  \hbox {\scriptsize phys}}^-$).
Thus we shall
work in the Hilbert space
${\cal H}_{s \,  \hbox {\scriptsize phys}}^+\otimes
{\overline {\cal H}}_{s \,  \hbox {\scriptsize phys}}^+\otimes
{\cal H}'\,\!_{-s \,  \hbox {\scriptsize phys}}^{-} \otimes
{\overline  {\cal H}}'\,\!_{-s \,  \hbox {\scriptsize phys}}^{-}$
(this  notation should be self explanatory).
The
BRST cohomology
selects  the states such that $L_0+L'_0$, and $\bar L_0+\bar L'_0$ have
eigenvalues equal to  one, which correspond to the vertex Eq.\ref{7.5}.
In the same way, as in the
weak coupling regime, this is  satisfied if $\varpi^2=\varpibp^2$,
and $\varpib^2=\varpihatbp^2$. We shall make  choices which are
consistent with Eq.\ref{7.6}
by letting $\varpi=-\varpihatp$ (or equivalently
$\varpihat=\varpip$), and $\varpib=\varpihatbp$ (or equivalently
$\varpihatb=-\varpibp$). Thus the spectrum of $\varpi$ is given by
\beqa
\varpi_{r,n} &=&-\left ({r\over 2-s} +n\right )(1-{\pi\over h}),\quad
\varpip_{r,n}=\left ({r\over 2-s} +n\right )(1-{h\over \pi}); \nnn
\varpib_{\rb, \nb}&=
&\left ({\rb\over 2-s} +\nb\right )(1-{\pi\over h}),\quad
\varpibp_{\rb, \nb}=\left ({\rb\over 2-s} +\nb\right )(1-{h\over \pi})
\label{7.7}.
\eeqa
The corresponding spins are
\beqa
J_{r,n}&=& -{r\over 2(2-s)}-{n+1\over 2},\quad
\Jhat_{r,n}=J'_{r,n}=\Jhat'_{r, n}= {r\over 2(2-s)}+{n-1\over 2}, \nnn
\Jhatb_{\rb,\nb}&=& -{\rb\over 2(2-s)}-{\nb+1\over 2},\quad
\Jb_{\rb,\nb}=\Jb'_{\rb,\nb}=\Jhatb'_{\rb, \nb}=
{\rb\over 2(2-s)}+{\nb-1\over 2}.
\label{7.8}
\eeqa
Clearly, one has $J_{r,n}=-\Jhat_{r,n}-1$,
$J'_{r,n}=\Jhat'_{r,n}$, and so on,  so that the spectrum of
highest-weights states (Eq.\ref{7.7}),
 and of vertex operators (Eq.\ref{7.6}),  are identical.

Our next point is the choice of $J$ and $\Jb$ in the vertex
operator Eq.\ref{7.5}. The basic
requirement is that these operators must   commute at equal $\tau$.
It follows from Eqs.\ref{brdchi-} and \ref{brdchi+} that
\beq
{\cal V}_{J'_1, \Jb'_1}(\sigma_1, \tau)
{\cal V}_{J'_2, \Jb'_2}(\sigma_2, \tau)
=e^{i\pi 2(2-s) (2J'_1J'_2-2\Jb'_1\Jb'_2)\epsilon(\sigma_1-\sigma_2) }
{\cal V}_{J'_2, \Jb'_2}(\sigma_2, \tau)
{\cal V}_{J'_1, \Jb'_1}(\sigma_1, \tau)
\label{7.9}.
\eeq
It is easy to see that the exponential factor becomes  equal to one
 if we impose
\beq
J'_i-\Jb'_i=\nu_i, \quad \nu_i \in Z.
\label{7.10}
\eeq
 Indeed
 it  becomes  equal to
$\exp \{- i \pi 2(2-s)(2J'_1\nu_2+2J'_2\nu_1)\} $.
It  follows from Eq.\ref{7.8} that $2(2-s) J'_i \in Z$,  which completes
the derivation.  Thus we shall consider that  $J'$
and $\Jb'$ in Eq.\ref{7.5}
are constrained to satisfy a condition of the type Eq.\ref{7.10}. Next
let us show that this is consistent with  the
requirement that the properties of  vertex operators
 be  symmetric between the three legs. Indeed, it follows from
the definitions  Eqs.\ref{chi-}, and \ref{chidef+} that, in
general, $<\varpi_{L'} | \chi_\pm ^{(J')} |\varpi_{K'}> \not= 0$ only for
$J'+K'-L'\in Z_+$. For the vertex Eq.\ref{7.5}, where $\Jb'-J'\in Z$,
and if  we consider  matrix elements of the type just written, with
${\overline K}' -K'\in Z$, we will also
have ${\overline L}'-L' \in Z$. Thus it is consistent
to assume that  conditions of the type Eq.\ref{7.10} hold for
the three legs of
the vertex operators.  Note that the stronger condition
${\overline K}'-K'=0$ is not
preserved since in Eq.\ref{7.5} the summations  over left and right
movers are independent. Thus  it would be inconsistent to consider
vertex operators with $J'=\Jb' $only.
Since we take $J'\not=\Jb'$, we have to discuss the monodromy
properties of our operators. In general, for
$V_{m\mhat}^{(J \Jhat)}$ they are given by
\beq
V_{m\mhat}^{(J \Jhat)}(\tau+i\sigma+2i \pi)
=e^{-ih\varpi^2/2} V_{m\mhat}^{(J \Jhat)}(\tau+i\sigma)
e^{ih\varpi^2/2}
\label{7.11}.
\eeq
It thus follows from Eqs.\ref{chi-}, and \ref{chidef+} that
$$
<\varpi_{J_{12}^{e \pm}}| \chi_{\pm}^{(J)}(\tau, \sigma+2 \pi)
|\varpi_{J_{2}^{e \pm}}>=
$$
\beq
e^{\mp i \pi  2 (2\mp s) (J_{12}-J_2)(J_{12}+J_2+1)}
<\varpi_{J_{12}^{e \pm}}| \chi_{\pm}^{(J)}(\tau, \sigma)
|\varpi_{J_{2}^{e \pm}}>
\label{7.12} .
\eeq
Consider the vertex for gravity, that is $\chi_{+}^{(J)} \chib_{+}^{(\Jb)}$.
One gets the factor
$$
e^{-i \pi 2 (2- s) [(J_{12}-J_2)(J_{12}+J_2+1) -
(\Jb_{12}-\Jb_2)(\Jb_{12}+\Jb_2+1)]}
{}.
$$
It follows from the conditions  $J_1+J_2-J_{12}\in Z_+$,
$\Jb_1+\Jb_2-\Jb_{12}\in Z_+$ of Eq.\ref{chidef+}, and from Eq.\ref{7.10},
that $J_2-J_{12} -(\Jb_2-\Jb_{12}) \in Z$. Since the values of $J$ are such
that $2(2-s)J$ is an integer, one easily deduces that the last exponent
is equal to $i \pi$ multiplied by an integer. Thus the gravity
part is at most multiplied by a minus sign when $\sigma\to \sigma+2 \pi$.
On the other hand, using the special choice of
$J$'s and $\Jb$'s   (Eqs.\ref{7.6}), one easily sees that
the matter part gives the same factor, so that
${\cal V}_{J \Jb}$ is invariant under  this $2\pi$ rotation as expected
physically.

 In the weak
coupling regime, the Liouville field, and
thus the vertex operators preserve\cite{G5}  the condition $J'-\Jb'=0$.
 This is not true for our vertex operator  in the strong
coupling regime.
Thus the transition through the $c=1$ barrier may be considered as related
with a
deconfinement of this  quantum  number.
\subsection{The three-point function}
Next, our aim is to compute the three-point functions of the form
\begin{equation}
\bigl <
\prod_{\ell=1}^3 {\cal V}_{J_\ell,\,\Jb_\ell}(\sigma_\ell,\,\tau_\ell)
 \bigr > =
{\cal C}_{1,2,3}\Bigr /
\Bigl (\prod_{k<l} \vert z_k-z_l \vert^2 \Bigr  ), \quad
z_\ell=e^{\tau_\ell+i\sigma_\ell}
\label{7.13}
\end{equation}
where ${\cal C}_{1,2,3}$ is the coupling  constant.
Applying the same reasoning\footnote{The left- and right-most operators
require some special treatment as in the weak coupling regime. Since it is
completely similar, we do not discuss this point here   again.}
 as in ref.\cite{G5}, one sees that
$${\cal C}_{1,2,3}=
<-\varpi_0, -\varpib_0, -\varpi'_0 , -\varpib'_0 \vert  {\cal V}_
{J_3,\,\Jb_3} \vert  -\varpi_3, -\varpib_3, -\varpi'_3 -\varpib'_3 >\times
$$
$$
 <-\varpi_3, -\varpib_3,  -\varpi'_3 -\varpib'_3 \vert    {\cal V}_
{J_2,\,\Jb_2} \vert  \varpi_1, \varpib_1,  \varpi'_1 \varpib'_1 >
\times
$$
\beq
< \varpi_1, \varpib_1,  \varpi'_1 \varpib'_1\vert   {\cal V}_
{J_1,\,\Jb_1} \vert  \varpi_0, \varpib_0, \varpi'_0 , \varpib'_0 >,
\label{7.14}
\eeq
where the $\varpi_i$'s are  associated with the spin $J_i$, and
of the type Eq.\ref{7.6}, Eq.\ref{7.7}.
 By definition,
the highest-weight matrix elements of the $V$ fields are  equal to
one or zero. It thus follows,
from Eqs.\ref{chidef-}, \ref{chidef+} and \ref{chidef+p}  that
$$
<\varpi_k, \varpib_k,  \varpi'_k \varpib'_k \vert    {\cal V}_
{J,\,\Jb}
\vert  \varpi_\ell, \varpib_\ell,  \varpi'_\ell \varpib'_\ell >=
(-1)^{(2-s)[ 2J_\ell(J+J_\ell-J_k)+ 2\Jb_\ell (\Jb+\Jb_\ell-\Jb_k)]}
\times
$$
\beq
(-1)^{(2-s)[
2J'_\ell(J'+J'_\ell-J'_k)+ 2(\Jb'_\ell+1) (\Jb'+\Jb'_\ell-\Jb'_k+1)]}
g_{\Je, \Jne{\ell} }^{\Jne{k} }\>  \gp_{\Jep, \Jnep{\ell} }^{\Jnep{k} }\>
\gb_{\Jeb, \Jneb{\ell} }^{\Jneb{k} }\>
\gbp_{\Jebp, \Jnebp{\ell} }^{\Jnebp{k} }
\label{7.15}
\eeq
where the spins are the appropriate effective ones,
provided the appropriate
differences between spins are integers (recall Eq.\ref{OneCond}).
Next, it is easy to see
that the first and the last matrix elements in Eq.\ref{7.14} involve
coupling constants with one vanishing spin (of the form $g_{\Je, 0}^{\Je}$
or $g_{\Je, -\Je-1-\pi/h}^{-1-\pi/h}$)  which are equal to one. The middle one,
gives
\beq
{\cal C}_{1,2,3}=
g_{\Jne{2} , \Jne{1} }^{-\Jne{3} -1-\pi/h}\>\>
\gp_{\Jnep{2} , \Jnep{1} }^{-\Jnep{3} -1-\pi/h'}\>\>
\gb_{\Jneb{2} , \Jneb{1} }^{-\Jneb{3} -1-\pi/h}\> \>
\gbp_{\Jnebp{2} , \Jnebp{1} }^{-\Jnebp{3} -1-\pi/h'}
\label{7.16}
\eeq
where  the effective spins take the form displayed on Eqs.\ref{7.6}.
These coupling constants are given\footnote{directly, or  after
continuation using  the symmetry $J\to -J-1$,}  by
formulae of the type Eq.\ref{ggen}, with
\beq
\left \{
\begin{array}{cc}
p\equiv J_1+J_2+J_3+1, & \phat\equiv \Jhat_1+\Jhat_2+\Jhat_3+1, \nnn
p'\equiv J'_1+J'_2+J'_3+1,& \phat'\equiv \Jhat'_1+\Jhat'_2+\Jhat'_3+1,\nnn
\pb \equiv J_1+J_2+J_3+1,&
\phatb \equiv \Jhatb_1+\Jhatb_2+\Jhatb_3+1, \nnn
\pb' \equiv \Jb'_1+\Jb'_2+\Jb'_3+1,&
\phatb' \equiv \Jhatb'_1+\Jhatb'_2+\Jhatb'_3+1.
\end{array} \right.
\label{7.17}
\eeq
Applying a reasoning similar to the one already given\cite{G5} for the
weak coupling regime, one easily sees that the three
sets  of $J$'s satisfy the same relations:
\beq
\begin{array}{cccc}
J_i=-\Jhat_i'-1,&  \Jhat_i=J'_i,& J_i=-\Jhat_i-1,&  J'_i=\Jhat'_i\nnn
\Jb_i=\Jhatb_i',&  \Jhatb_i=-\Jb'_i-1,&
\Jhatb_i=-\Jb_i-1,&  \Jb'_i=\Jhatb'_i.
\end{array}
\label{7.18}
\eeq

Next we determine the product of the first two terms
of Eq.\ref{7.16} --- the last two
are similar. According to Eq.\ref{ggen}, we have
\beq
g_{\Jne{2} , \Jne{1} }^{-\Jne{3} -1-\pi/h}\>\>
\gp_{\Jnep{2} , \Jnep{1} }^{-\Jnep{3} -1-\pi/h'}=
\left ({i\over 2}\right )^{p+\phat+p'+\phat'}
{\prod_{k=1}^3 H_{p\phat}(\varpi_k) H'_{p'\phat'}(\varpi'_k)
\over H_{p\phat}(\varpi_{p/2, \phat/2})
H'_{p'\phat'}(\varpi'_{p'/2, \phat'/2})}
\label{7.19} .
\eeq
The p-parameters are related by
\beq
p'=\phat',\quad  p=-\phat-1,\quad
p=-\phat'-1,\quad  \phat=p'
\label{7.20}
\eeq
the first two relations
 are specific to the strong coupling regime, while the last two
are not.
The calculation we are going to perform will actually not make use of the
former  so that it also applies to the weakly coupled case.  First consider
the numerator in  the r.h.s. of Eq.\ref{7.19}. Each term is of the same type.
So we compute
$$
H_{p\phat}(\varpi) H'_{p'\phat'}(\varpi') =
\prod_{r=1}^p \sqrt{F(\varpihat-rh/\pi)}
\prod_{\rhat'=1}^{\phat'} \sqrt{F(\varpi'-\rhat'\pi/h')}
\times
$$
\beq
{\prod_{\rhat=1}^{\phat} \sqrt{F(\varpi-\rhat \pi/h)}
\prod_{r'=1}^{p'} \sqrt{F(\varpihat'-r'h'/\pi)} \over
\prod_{r=1}^p \prod_{\rhat=1}^{\phat}
\left ( (\varpi-r) \sqrt {h/\pi} -\rhat \sqrt{ \pi /h}\right )
\prod_{r'=1}^{p'} \prod_{\rhat'=1}^{\phat'}
\left ( (\varpi'-r') \sqrt {h'/\pi} -\rhat' \sqrt{ \pi /h'}
\right) }
\label{7.21}.
\eeq
Each pair of factors is seen to  simplify  tremendously, using the fact that
$\varpi=-\varpihat'$, and $\varpihat=\varpi'$. One finds
$$
\prod_{r=1}^p \sqrt{F(\varpihat-rh/\pi)}
\prod_{\rhat'=1}^{\phat'} \sqrt{F(\varpi'-\rhat'\pi/h')}=
{\prod_{r=1}^p \sqrt{F(\varpihat-rh/\pi)}\over
\prod_{r=1}^{p+1} \sqrt{F(\varpihat-(r-1)h/\pi)}}
={1\over \sqrt{F(\varpihat)}},
$$
where we use the general substitution rule Eq.\ref{prodneg}
 that follows from the symmetry $J\to -J-1$. On the other hand,
$$
\prod_{\rhat=1}^{\phat} \sqrt{F(\varpi-\rhat \pi/h)}
\prod_{r'=1}^{p'} \sqrt{F(\varpihat'-r'h'/\pi)}=
\prod_{\rhat=1}^{\phat}
\sqrt{F(\varpi-\rhat \pi/h)}
\sqrt{F(-\varpi+\rhat \pi/h)}
$$
$$=\prod_{\rhat=1}^{\phat}  i \Bigl /(\varpi-\rhat \pi/h).
$$
Combining with the denominator, we finally derive
\beq
H_{p\phat}(\varpi) H'_{p'\phat'}(\varpi') =
(\pi /h)^{\phat/2} e^{i\pi \phat (p+1)/2} e^{i\pi \phat /2}
\Bigl / \sqrt{ F(\varpihat)}
\label{7.22}.
\eeq
The calculation of the denominator of Eq.\ref{7.14} is not quite  the same
 since the
relation between $\varpi_{p/2, \phat/2}$ and $\varpi'_{p'/2, \phat'/2}$
differs  from the one between $\varpi_k$ and $\varpi'_k$.
Applying the same method, one finds for the two  pairs of factors
$$
\prod_{r=1}^p \sqrt{F\left(\phat+1+(p-r+1)h/\pi\right)}
\prod_{\rhat'=1}^{\phat'} \sqrt{F\left(\phat+1-(p+\rhat')\pi/h'\right)}=
{1\over \sqrt{F(\phat+1)}},
$$
$$
\prod_{\rhat=1}^{\phat} \sqrt{F\left(p+1+(\phat-\rhat+1)\pi/h\right)}
\prod_{r'=1}^{p'} \sqrt{F\left(-p-(p'-r'+1)\pi/h\right)}=
1.
$$
One easily sees that  the denominator is equal to
$ \exp [i\pi \phat (p+1)/2] (h/\pi )^{\phat/2} /\phat  !$.
Using the fact that  $F(\phat+1)=
 \phat !^2\sin (\pi(\phat+1))/\pi$, one finally finds
\beq
H_{p\phat}(\varpi_{p/2, \phat/2})
H'_{p'\phat'}(\varpi'_{p'/2, \phat'/2})= e^{-i\pi (\phat+1)/2}
e^{-i\pi\phat  (p+1)/2}
\left ({h\over\pi}\right) ^{-\phat/2}\sqrt{\pi}
\label{7.23}.
\eeq
It immediately follows from Eq.\ref{7.17} that any term of the form
$(\beta)^{\phat}$ may be written as $\beta \prod_{k=1}^3 (\beta)^{\Jhat_k}$.
Using this remark, to simplify Eq.\ref{7.22} and \ref{7.23}, we
finally  derive the following
expression of the l.h.s. of Eq.\ref{7.19}
\beq
g_{\Jne{2} , \Jne{1} }^{-\Jne{3} -1-\pi/h}\>\>
\gp_{\Jnep{2} , \Jnep{1} }^{-\Jnep{3} -1-\pi/h'}=
a (-1)^{p\phat} \prod_{k=1}^3
{(b)^{\Jhat_k}\over \sqrt{F\left (2\Jhat_k+1+(2J_k+1)h/\pi\right)}}
\label{7.24}
\eeq
where $a$ and $b$ are independent from the $J$'s.

Concerning the second factor in Eq.\ref{7.15},
the calculation is essentially the same. The only difference is the
choice of $\Jb$, and $\Jhatb$. One may see   that it amounts to
exchanging  the r\^oles of unhatted and hatted quantities. Combining
everything together one gets
$$
{\cal C}_{1,2,3}= a{\overline  a} (-1)^{p}
(-1)^{(2-s)2(J_1p+\Jb_1 \pb +J_1'p'+(\Jb_1'+1)(\pb'+1))}
\times
$$
\beq
\prod_{k=1}^3 { (b)^{\Jhat_k} ({\overline b})^{\Jb_k}
\over \sqrt{F\left (2\Jhat_k+1+(2J_k+1)h/\pi\right)}
\sqrt{F\left (2\Jb_k+1+(2\Jhatb_k+1)\pi/h\right)}}
\label{7.25}.
\eeq
The $-1$ factor may be simplified. Using Eqs.\ref{7.18}, \ref{7.20}
one easily sees that
$$
(-1)^{(2-s)2[J_1p+\Jb_1 \pb +J_1'p'+(\Jb_1'+1) (\pb'+1)]}=
(-1)^{(2-s)(J_1+ \Jb_1)} =1.
$$
 The final result is
\beq
{\cal C}_{1,2,3} =
a{\overline  a} \prod_{k=1}^3 { (b)^{J'_k} ({\overline b})^{\Jb'k}
 \over
\sqrt{F\left ((2J'_k+1)(1-{\displaystyle {h\over \pi}})\right)}
\sqrt{F\left ((2\Jb'_k+1)(1-{\displaystyle{\pi\over h} }) \right)}}
\label{7.26}.
\eeq
Since $h$ is not real, this three-point coupling is complex.
However, taking complex conjugate simply exchanges  $J'_k$ with $\Jb'_k$.
Thus left and right movers are exchanged, which makes sense physically.

\section{OUTLOOK}
\markboth{8. Outlook}{8. Outlook}
There are several interesting physical points to make about our results.
In particular, the vertex ${\cal V}_{0,0}$ seems to define a
cosmological term which differs drastically from the one (the Liouville
exponential) which is relevent for the weak-coupling regime. Its
study should throw light on the nature of the ``c=1'' barrier.
The associated string susceptibility seems computable.
The result is real,  while the
weak-coupling formula is complex in the strong coupling case.
{}From the viewpoint of conformal theories, the cosmological
term is the  marginal operator that takes us away from free-field theory.
Thus at $c=1$ a new marginal operator replaces the standard one,
and this is why the theory for $c>1$ looks so different.
We have seen that the barrier
 seems to be related to a deconfinement of chirality.
This is also clear on  the expression of
${\cal V}_{0,0}$: it  corresponds to a metric
tensor which is a simple product of one analytic function
 by  its   anti-analytic  counterpart.
  In a way, the surface becomes degenerate.
We shall return to these points in a separate article.
Another remark  is that
the present new topological models may be   simple enough
so that their n-point functions are  derivable  in closed  form.

The quantum group technology we have developed, is clearly
interesting in itself. It should be helpful,
 to make progress for the
Liouville string theories  in full-fledged space-times.
In the weak coupling regime, going to continuous $J$ also seems
a key step. For instance, it allows  to define the
Liouville field itself\cite{GS3}.

The present method should be extendable to the $N=1$ super-Liouville
theory, making use of the quantum group structure exhibited in
ref.\cite{GR2}. This will be useful to study the Liouville
superstrings\cite{BG}, which are very interesting physically.
In particular the five-dimensional model seems  related with
the supersymmetric string solutions recently put
forward in ref.\cite{AFK}.

There are clearly many more relevent comments. We leave them  for future
publications.
\bigskip

\noindent {\large \bf Acknowledgements}

We are indebted  to Jens Schnittger for repeated discussions
that were very useful. This work was supported in part by the E.U.
network ``Capital Humain et Mobilit\'e'', contract \# CHRXCT920035.

\appendix
\section{Appendix: Relation with the earlier definition}
\markboth{Appendix}{Appendix}
In this appendix, we compare our general expressions for the $\chi$ fields
with the earlier one of ref.\cite{G3}.
For this, it is necessary
to use the fact that
 braiding and fusing equations are invariant under
the following, which we  call gauge transformations:
\beq
V_{m,\, \mhat}^{(J, \Jhat )}\to
S(\varpi) V_{m,\, \mhat}^{(J, \Jhat)} S^{-1} (\varpi),
\quad  g_{\Jgen1 \Jgen2 }^{\Jgen{12} } \to {S(\varpi_{\Jgen{12} })
\over S(\varpi_{\Jgen{2} })} g_{\Jgen1 \Jgen2 }^{\Jgen{12} },
\label{2.11}
\eeq
where $S$ is an arbitrary function of $\varpi$, independent
from the
other mode of the underlying B\"acklund free field\footnote{This idea
was already used in ref.\cite{GS3} to show that the different  operator
quantizations of Liouville theory are actually related by transformations
of this type.}. One may show, using\footnote{For this it is
convenient to choose $g_0=h/\pi$, instead of $g_0=2\pi$ as was done
in refs.\cite{CGR1}, \cite{CGR2}.}  Eq.3.28 of
ref.\cite{CGR1},  that the relation between $\psi$ and $V$
fields may be cast under the
form (we do not specify the indices $\underline K$, $\underline L$,
$\underline M$ of $g$ for brevity)
\beq
\psi_{m,\, -m}^{(J, J)}\propto \rho(\varpi)
{1\over \sqrt{ C_m^{(J)}(\varpi) \Chat _{-m}^{(J)}(\varpihat)}}
g_{\underline K, \underline L }^{\underline M} \>
V_{m,\, -m}^{(J, J)} \> \rho^{-1}(\varpi).
\label{2.12}
\eeq
where
\beq
\rho(\varpi) = {1
\over \sqrt{ \Gamma[\varpi h/\pi]
\Gamma[ \varpihat \pi/h] {\sqrt {\varpi \varpihat}}}}
\label{6}.
\eeq
Substituting into Eq.\ref{2.9}, one finds
\beq
\chi_{- \hbox{\scriptsize  G.}}^{(J)} \propto
\rho(\varpi) \left (
\sum_{m=-J}^J e^{im [\hhat \varpihat -h \varpi+(h-\hhat)/2]}
\sqrt{ C_m^{(J)}(\varpi) \over \Chat _{-m}^{J)}(\varpihat)}
g_{\underline K, \underline L }^{\underline M}
 V_{m,\, -m}^{(J, J)}\right )  \rho^{-1}(\varpi).
\label{2.13}
\eeq
On the other hand, one may check that
\beq
\sqrt{ C_m^{(J)}(\varpi) \over \Chat _{-m}^{J)}(\varpihat)}=
\propto e^{iJ [\hhat \varpihat -h \varpi]} e^{i\pi s[Jm+m^2/2]}.
\label{phase}
\eeq
We dropped factors that only depend upon $J$, and factors of the
form $\exp (2\alpha m)$, $\alpha$ independent from $J$ and $\varpi$. The
latter factors are pure gauges in the sense of Eq.\ref{2.11}. Indeed
one has
$$
e^{2\alpha m} V_{m,\, -m}^{(J, J)}=
e^{-\alpha \varpi/(1-\pi/h)} V_{m,\, -m}^{(J, J)}
e^{\alpha \varpi/(1-\pi/h)}
$$
as
is easily verified. The terms that only depend upon $J$
 only change the overall normalization
which could not be seen in ref.\cite{G3} where the coupling constants were
left undetermined. On the other hand, we have
$$
e^{im (h-\hhat)/2}V_{m,\, -m}^{(J, J)}= e^{-ih(1+\pi/h) \varpi/4}
V_{m,\, -m}^{(J, J)} e^{ih(1+\pi/h) \varpi/4}
$$
so that the factor $\exp{im (h-\hhat)/2}$ in Eq.\ref{2.13} may be
absorbed by a change of $\rho(\varpi)$. Collecting all the factors, one
finally finds
\beq
\chi_{- \hbox{\scriptsize  G.}}^{(J)} \propto
\widetilde \rho(\varpi) \left (
\sum_{m=-J}^J e^{i \pi s [2J_{12}  (J+m) -(J+m)(J+m+1)/2]}
g_{\underline K, \underline L }^{\underline M}
 V_{m,\, -m}^{(J, J)}\right ){\cal P}_{J_{12}}  \widetilde \rho^{-1}(\varpi).
\label{2.13f}
\eeq
The symbol $\widetilde \rho$ stands for the modified gauge transformation
operator. Up to this gauge transformation, this expression
 coincides with our general formula Eq.\ref{chi-} for
half-integer $J$'s. Thus the normalizations  of the present article are
deduced from the previous one by a gauge transformation of the type
Eq.\ref{2.11}. Note that $\chi_{- \hbox{\scriptsize  G.}}^{(J)}$ is not
really equivalent to the one we have  introduced, since it was defined in
${\cal H}^+_{s \,  \hbox {\scriptsize phys}} $ instead of
${\cal H}^-_{s \,  \hbox {\scriptsize phys}} $. The expression
just given is unsensitive to this modification.

\end{document}